\documentstyle[12pt]{article}
\renewcommand{\v}{${\cal V}ir\hspace{-.03in}_{p,q}\,$}
\newcommand{\Winf}{{\cal W}_{1+\infty}}

\newcommand\SC{X} 
\newcommand\QQ{P_-}
\newcommand\PP{P_+}
\newcommand\RR{r} 
\newcommand\SSS{s} 
\newcommand\cF[2]{{\cal F}_{{#2,#1}}}
\newcommand\cL[2]{{\cal L}_{#2,#1}}
\newcommand\DEL[2]{h_{#2,#1}}
\newcommand\EPS[2]{{\varepsilon}_{#2,#1}}

\newcommand\QINT[2]{#1^{#2\over2}-#1^{-{#2\over2}} }
\newcommand\PINT[2]{#1^{#2\over2}+#1^{-{#2\over2}} }

\newcommand{\La}{\Lambda}

\newcommand{\bN}{{\bf N}}
\newcommand{\bZ}{{\bf Z}}
\newcommand{\bC}{{\bf C}}

\newcommand\BRS[1]
{\,{\displaystyle\mathop{-\!\!\!-\!\!\!-\!\!\!-\!\!\!\longrightarrow}^{#1}}\,} 

\newcommand{\be}{\begin{equation}}     
\newcommand{\ee}{\end{equation}}
\newcommand{\ba}{\begin{eqnarray}}     
\newcommand{\ea}{\end{eqnarray}}
\renewcommand{\&}{&\!\!\!\!\!\!\!\! &}

\newcommand{\nn}{\nonumber}

\newcommand{\eq}[1]{(\ref{#1})}

\newcommand{\tpsi}{J}   
\renewcommand{\varphi}{\widehat J}

\newcommand{\llra}
{\!\!\!-\!\!\!-\!\!\!-\!\!\!-\!\!\!-\!\!\!-\!\!\!\longrightarrow}
\newcommand{\llla}
{\longleftarrow\!\!\!-\!\!\!-\!\!\!-\!\!\!-\!\!\!-\!\!\!-}
\newcommand{\ps}[1]{\sum^{\infty}_{{#1}=1}}
\newcommand{\pzs}[1]{\sum^{\infty}_{{#1}=0}}

\newcommand{\bea}{\begin{eqnarray}}
\newcommand{\eea}{\end{eqnarray}}
\newcommand{\kv}[1]{{|#1 \rangle}}
\newcommand{\ket}[1]{{|#1 \rangle}}
\newcommand{\bv}[1]{{\langle #1 |}}

\renewcommand{\v}{${\cal V}ir\hspace{-.03in}_{p,q}\,$}

\newcommand{\ignore}[1]{}
\newcommand{\n}{\nonumber\\}

\renewcommand{\i}{\sqrt{-1}}
\begin{document}
%
\renewcommand{\thefootnote}{\fnsymbol{footnote}}
\font\csc=cmcsc10 scaled\magstep1
{\baselineskip=14pt
 \rightline{
 \vbox{
       \hbox{EFI-96-44}
       \hbox{DPSU-96-18}
       \hbox{UT-764}
       \hbox{December 1996}
}}}
~~\vskip 25mm
\begin{center}
{\large\bf
Virasoro-type Symmetries in
Solvable Models}\footnote[2]{Based on 
talks presented by H.A. and J.S.
at the Nankai-CRM joint meeting on the quantum deformed 
Virasoro algebra, Tianjin, China, August 19-24, 1996. 
To appear in the CRM series in mathematical physics, Springer Verlag.
}

\vspace{15mm}
{\csc Hidetoshi AWATA}\footnote[1]{JSPS fellow}$^{1}$,
{\csc Harunobu KUBO}$^{*2}$,
\\ 
\vskip.05in
{\csc Satoru ODAKE}$\,{}^3$ and
{\csc Jun'ichi SHIRAISHI}$\,{}^4$
{\baselineskip=15pt
\it\vskip.35in 
\setcounter{footnote}{0}\renewcommand{\thefootnote}{\arabic{footnote}}
\footnote{e-mail address : awata@rainbow.uchicago.edu}
James Frank Institute and Enrico Fermi Institute,
University of Chicago,\\
5640 S. Ellis Ave., Chicago, IL 60637, U.S.A.
\vskip.1in 
\footnote{e-mail address : kubo@hep-th.phys.s.u-tokyo.ac.jp}
Department of Physics, Faculty of Science \\
University of Tokyo, Tokyo 113, Japan \\
\vskip.1in 
\footnote{e-mail address : odake@azusa.shinshu-u.ac.jp
}
Department of Physics, Faculty of Science \\
Shinshu University, Matsumoto 390, Japan\\
\vskip.1in 
\footnote{e-mail address : shiraish@momo.issp.u-tokyo.ac.jp}
Institute for Solid State Physics, \\
University of Tokyo, Tokyo 106, Japan \\
}
\end{center}
 
\vspace{10mm}
\begin{abstract}

Virasoro-type symmetries and 
their roles in solvable models are reviewed. 
These symmetries are described by 
the two-parameter Virasoro-type 
algebra ${\cal V}ir_{p,q}$
by choosing the parameters $p$ and $q$ suitably.

\end{abstract}

\vspace{7mm}
\renewcommand{\thefootnote}{\arabic{footnote}}
\vfill
\newpage
%
%
\section{Introduction}
Integrable models may have very huge symmetries, that 
help us to study various behaviors of the systems.
For some well investigated models, we are able to 
calculate correlation functions of observables 
by using representation theories of the symmetries, that
have been fruitfully studied within the language of 
infinite dimensional Lie algebras and their suitable
deformation theories. We now have a lot of
examples of these symmetry algebras and their applications 
to many problems.
What is the natural candidate for that symmetry which is 
present in majority of the integrable models?
In other words, what is the ``universal'' symmetry among all these?

The $1+1$-dimensional conformal field theories (CFT's) 
\cite{rBPZ} describe the 
universality classes of massless field theories in $1+1$-dimensions.
It is the guideline given in the celebrated paper
\cite{rBPZ} 
that all the CFT's must be 
regarded as representations of the ``Virasoro algebra''
regardless of the details of the models.
Namely, the Virasoro algebra is the universal one in CFT.
Indeed, through the Sugawara construction \cite{rKZ,rGO},
the Virasoro algebra exists in any Kac-Moody algebra.
These models are successfully 
applied to the critical phenomena of 
two-dimensional classical statistical models,  
the low temperature behavior of one-dimensional electron systems and so on.
In any sense, CFT is definitely quite well understood among 
other interacting field theories, 
because we have the infinite dimensional conformal symmetry.

We are gradually understanding the universal symmetry arising 
in off-critical models: massive integrable field
theories in 1+1 dimension (sine-Gordon model etc.) \cite{rL}, 
one-dimensional quantum spin chain systems ($XYZ$ model etc.) \cite{rL},
two-dimensional solvable lattice 
models (ABF models etc.) \cite{rLP1,rLP2,rAJMP},
deformations of the KdV hierarchy and 
discretized soliton equations \cite{rFR,rFr},
Calogero-Sutherland(CS)-type quantum mechanical 
models \cite{rAMOS,rAOS} and so on.
It had been recognized that nontrivial
Virasoro-type symmetries exist in these off-critical theories;
in \cite{rLP1}, the existence of two-parameter
Virasoro-type symmetry was conjectured, and 
a one-parameter Virasoro-type Poisson structure was
found in \cite{rFR}.
It was in the CS-type quantum 
mechanical model, that 
we finally obtained the definition and an 
explicit construction of the two-parameter
Virasoro-type symmetry which we call \v \cite{rSKAO}.
This is already extended to the case of ${\cal W}$-algebra \cite{rFFr,rAKOS}.
It is really astonishing that all the Virasoro-type algebras 
related with the off-critical integrable models are 
obtained by taking suitable limits from \v.
However,
we have not fully understand 
the meaning of ``universality'' played by this new Virasoro-type
symmetry \v, so far. 
The algebra \v is one of the simplest example of 
elliptic algebras \cite{rFIJKMY,rFFr}, since we have elliptic theta-functions
in the operator product expansion (OPE) formulas.
We strongly hope that this Virasoro-type algebra 
will be constructed in a canonical way from 
the elliptic algebra ${\cal A}_{q,p}$ \cite{rFIJKMY}\footnote{
The relations between the parameters in ${\cal A}_{q,p}$ and \v 
are $q_{\!\!~_{{\cal V}ir}}=p_{\!\!\!~_{{\cal A}}}$ and
$p_{\!\!~_{{\cal V}ir}}={q_{\!\!\!~_{{\cal A}}}}^2$.}
through a Sugawara-type construction and 
that will give us a clue for the total understanding of \v.

In this review, we will explain how this two parameter 
Virasoro-type algebra \v
arose in the CS-type model, and
another aim is to accumulate
as many problems and applications of \v as possible.
This paper is organized as follows.
In Section 2, we study basic results of the Virasoro-type algebras 
starting from the definition of the two-parameter Virasoro-type algebra \v. 
In Section 3, a Heisenberg realization of the Virasoro-type algebra and 
its applications to the CS-type models are presented.
Based on this realization, we discuss
representation theories, futher applications and 
relations with other several models, in Section 4.
Section 5 is devoted to summary and comments.

%
%
\section{Quantum deformed Virasoro algebra \v }
In this section we examine some of the fundamental properties of 
the Virasoro-type algebra
\v \cite{rSKAO} 
which can be directly derived from the defining relation. 
A Heisenberg realization and its application to 
various problems are studied in the next sections.
\subsection{definition of \v}
Let $p$ be a generic complex parameter with $|p| < 1$.
Let us consider an associative algebra generated by
$\{T_n|n\in \bf{Z}\}$ with the relation
$$
f(w/z)T(z)T(w)-T(w)T(z)f(z/w)
= {\rm const.}\left[
       \delta \Bigl(\frac{pw}{z}\Bigr)-
       \delta \Bigl(\frac{p^{-1}w}{z}\Bigr)\right],
$$
where $T(z)=\sum_{n\in\bf{Z}}T_n z^{-n}$,
$\delta(x)=\sum_{n \in {\bf Z}}x^n$ and 
$f(z)=\pzs{l}f_l z^l$ is a structure function. We show that the 
commutativity of the diagram (Yang-Baxter equation for $T(z)$)
\begin{eqnarray}
\begin{array}{ccc}
 T(x)T(y)T(z) &{\displaystyle \mathop{\llra}^{f(z/y)\times} } &
T(x)T(z)T(y) f(y/z)    \\
& & + T(x)\; \delta{\rm-function}\\
{\scriptstyle f(y/x)\times} \Biggl\downarrow & & 
{\scriptstyle f(z/x)\times} \Biggl\downarrow \\
& & \\
T(y)T(x)T(z) f(x/y)  & & 
T(z)T(x)T(y) f(y/z) f(x/z)\\
+ T(z)\; \delta{\rm-function} & & + T(x)\; \delta{\rm-function}\\
& & + T(y)\; \delta{\rm-function}\\
{\scriptstyle f(z/x)\times} \Biggl\downarrow & & 
{\scriptstyle f(y/x)\times} \Biggl\downarrow \\
& & \\
T(y)T(z)T(x) f(x/y) f(x/z) &{\displaystyle \mathop{\llra}^{f(z/y)\times} } & 
T(z)T(y)T(x) f(x/y) f(x/z)f(y/z)\\
+ T(z)\; \delta{\rm-function} & & + T(z)\; \delta{\rm-function}\\
+ T(y)\; \delta{\rm-function} & & + T(y)\; \delta{\rm-function} \\
& & + T(x)\; \delta{\rm-function} 
\end{array}\label{YB-for-T}
\end{eqnarray}
determines this 
structure function $f(z)$ completely. 
Here the terms denoted by ``$\delta$-function'' 
mean some combinations of 
the $\delta$-functions and the structure functions.
The commutativity of this diagram means
\begin{eqnarray}
0&=&T(x) \Biggl( \delta(pz/y) g(x/z)-\delta(py/z) g(x/y) \Biggr) \nonumber\\
 &+&T(y) \Biggl( \delta(px/z) g(y/x)-\delta(pz/x) g(y/z) \Biggr) \label{asso}\\
 &+&T(z) \Biggl( \delta(py/x) g(z/y)-\delta(px/y) g(z/x) \Biggr), \nonumber
\end{eqnarray}
where
$g(x) \equiv f(x)f(x/p) - f(1/x) f(p/x)$.
Note that $g(x)=-g(p/x)$.
It is an interesting exercise to see that the general solution to the 
eq.\ (\ref{asso}) is 
$g(x)= c_1 \Bigl(\delta(x/p)-\delta(x) \Bigr)$,
where $c_1$ is a constant. 
Hereafter, we just set $c_1=1$,
since there is no loss of generality.
Note that the Yang-Baxter equation for $T(z)$ is ``not'' 
trivially satisfied even if the current $T(z)$ has
no spin degrees of freedom,
because we have the $\delta$-function term in the 
relation (\ref{e:a1.2}).
{}From this we have
\begin{eqnarray*}
f(x)f(x p)=  \alpha + \sum_{n=1}^\infty (1-p^{n})x^n ,
\end{eqnarray*}
where $\alpha$ is a constant.
It should be noted that this is the place where one more parameter comes in.
If we parameterize $\alpha$ by introducing another parameter $q$ as
$$
\alpha={1-p \over (1-q)(1-t^{-1})}, \qquad\quad t=qp^{-1},
$$
we have the difference equation 
$$
f(x)f(x p)=\alpha { (1-qx)(1-t^{-1} x) \over (1-x)(1-p x)}.
$$
This can be solved as
\begin{equation}
f(x)=\exp \Biggl\{ \sum_{n=1}^\infty 
{ (1-q^n)(1-t^{-n}) \over 1+p^n} { x^n \over n} \Biggr\}. \label{structure}
\end{equation}
We arrive at the definition of the quantum deformed Virasoro algebra
\v \cite{rSKAO}.

\proclaim Definition 1.
Let $p$ and $q$ be complex parameters with
the conditions $|p|<1$ and $|q|<1$, and set $t=qp^{-1}$.
The associative algebra \v is generated by the current
$T(z)=\sum_{n \in {\bf Z}} T_n z^{-n}$ satisfying the relation
\be
f(w/z)T(z)T(w)-T(w)T(z)f(z/w)
= -\frac{(1-q)(1-t^{-1})}{1-p} \left[
       \delta \Bigl(\frac{pw}{z}\Bigr)-
       \delta \Bigl(\frac{w}{pz}\Bigr)\right],
\label{e:a1.2}
\ee
with the structure function (\ref{structure}).

\noindent Note that the constant factor in the R.H.S. is so chosen 
that our Heisenberg realization of this algebra becomes simple.

We now have the associative algebra \v 
which exists in a nontrivial way,
since we have so chosen the structure function $f(z)$
that 
the Yang-Baxter equation for the \v current
(\ref{YB-for-T}) will not give us any more 
relations than the quadratic relation (\ref{e:a1.2}) for $T(z)$.

The defining relation (\ref{e:a1.2}) can be written in terms of $T_n$ as 
\be
[T_n \, , \, T_m]=-\ps{l}f_l\left(T_{n-l}T_{m+l}-T_{m-l}T_{n+l}\right)
-\frac{(1-q)(1-t^{-1})}{1-p}(p^{n}-p^{-n})\delta_{m+n,0},
\label{e:a1}
\ee
where $[A,B]=AB-BA$.

The relation \eq{e:a1.2} is invariant
under the transformations
\begin{eqnarray}
{\rm (I)}&& \qquad\qquad T_n \to -T_n, 
\label{e:a1.3} \\
{\rm (II)}&& \qquad\qquad (q,t)\to (q^{-1},t^{-1}),
\label{e:aa1} \\
{\rm (III)} &&\qquad\qquad q \leftrightarrow t. \label{e:aa2}
\end{eqnarray}
\par

In what follows, 
we will frequently use the notation
\begin{equation}
t=qp^{-1}=q^\beta.
\end{equation}
This parameter $\beta$ plays the role of the coupling constant of
the Calogero-Sutherland model. See Section \ref{secCS}.

\subsection{special limits of \v}
Here we study some special limits of \v, which 
explains the connections among known examples of the 
Virasoro-type algebras.

\subsubsection{limit of $q\rightarrow 1$: ordinary Virasoro algebra}
Let us study the limit $q\to 1$ ($\beta$: fixed) by
parameterizing  $q=e^{h}$. 
Suppose that $T(z)$ has the following expansion in $h$
\be
T(z)=2+\beta  \left( z^2 L(z)+\frac{(1-\beta)^2}{4\beta} \right)h^2
           + T^{(2)}(z)h^4 +\cdots.
\label{pe:a7}
\ee
This expansion is consistent with the invariance under transformation 
\eq{e:aa1}.
The defining relation \eq{e:a1.2} gives us the well known relations for
the ordinary Virasoro current
$ L(z)=\sum_{n \in {\bf Z}}L_{n}z^{-n-2}$, namely
\be
[L_n , L_m]=(n-m)L_{n+m}+\frac{c}{12}(n^3-n)\delta_{n+m,0},
\label{e:a8}
\ee
where the central charge $c$ is
\be
c=1-\frac{6(1-\beta)^2}{\beta}.
\label{e:a9}
\ee
This relation between  the central charge of
the Virasoro algebra and the coupling constant
of CS model is discussed in Section \ref{secCS}.

\subsubsection{Frenkel-Reshetikhin's $q$-Virasoro algebra
($\beta\rightarrow 0$)}
Let us consider the limit $\beta\to 0$ ($q$: fixed).
In this limit, we obtain the classical $q$-Virasoro algebra
found by Frenkel and Reshetikhin \cite{rFR}. Their algebra is 
the first one among many $q$-deformed Virasoro algebras,
which was obtained through the study of the quantum affine algebra
$U_q(\widehat{sl}_2)$ \cite{rAOS2}.
Other nice features of their algebra are
that it is no more a Lie algebra but a quadratic algebra 
which resembles the relation
$\{L(u),L(v)\}=[r(u-v),L(u)L(v)]$ 
in the quantum inverse scattering method,
and there exists a mysterious 
resemblance between their bosonic realization and the 
Baxter's
dressed vacuum form in the Bethe ansatz method. 
See Section \ref{secSYMMETRIC}.

In this limit, $T_n$'s become commutative. 
However, we can define the Poisson bracket structure
defined by $\{\;,\;\}_{\rm P.B.}=-\lim_{\beta\rightarrow 0}
[\;,\;]/(\beta \ln q )$.
Thus we have
\begin{eqnarray}
\{T_n,T_m\}_{\rm P.B.}=
\sum_{l\in {\bf Z}}{1-q^l \over 1+q^l} T_{n-l}T_{m+l}
+ (q^n-q^{-n}) \delta_{n+m,0}, \label{qvirFR}
\end{eqnarray}
which is the relations for the classical $q$-Virasoro algebra \cite{rFR}.

In the paper \cite{rFr}, deformations of the KdV hierarchy
is studied. The $N$-th Korteweg-de Vries (KdV) hierarchy
is a bihamiltonian integrable system with the 
$N$-th order differential operators.
They are deformed to the $q$-shift operators
$$
 D^N- t_1(z) D^{N-1}+\cdots +(-1)^{N-1} t_{N-2}(z) D
+(-1)^N t_{N-1}(z),
$$
where $D\cdot f(z)=f(qz)$.
It was shown there that the deformed bihamiltonian structure
is given by the Poisson bracket for the 
$q$-${\cal W}$ algebra of Frenkel and Reshetikhin \cite{rFR}.
\proclaim Proposition 1. \hspace{-2mm}\cite{rFR}
In the case of $N=2$, the bihamiltonian structure is 
given by
\begin{eqnarray}
\{t_i(z),t_j(w)\}_1&=&\delta\left( {wq\over z}\right)-
\delta\left( {w\over zq}\right),\\
\{t_i(z),t_j(w)\}_2&=&\sum_{m \in{\bf Z}} \left( {w\over z}\right)^m
{1-q^m \over 1+q^m} t(z)t(w)+
\delta\left( {wq\over z}\right)-
\delta\left( {w\over zq}\right).
\end{eqnarray}

\noindent
These Poisson brackets coincide with that of the 
$q$-Virasoro algebra of Frenkel and Reshetikhin (\ref{qvirFR})
with two different choices of the structure function $f(z)$:
{\it i.e.}, $f(z)=1$ and $f(z)$ given by (\ref{structure})
with $\beta\rightarrow 0$.

\subsubsection{Zamolodchikov-Faddeev algebra of sine-Gordon and XYZ models}
The $q$-Virasoro algebra can be interpreted as 
the Zamolodchikov-Faddeev (ZF) algebra satisfied by a 
particle excitation operator.
The structure function $f(x)$ determined by the associativity
relates with 
the $S$ matrix characterized by the factorization property.

First, we can rewrite the defining relation of the 
$q$-Virasoro algebra (\ref{e:a1.2})
as follows;
\be\label{eq:ZFXYZ}
T(z_1) T(z_2) 
= 
  S\left( {z_1\over z_2}\right) T(z_2) T(z_1)
+ C\left(  \delta\left( {z_1\over pz_2}\right) 
+ \delta\left( {pz_1\over z_2}\right) \right), 
\ee
with
\ba
S(z) 
&\!\!\!=\!\!\!&
{f(z) \over f(z^{-1})}=
 {\vartheta_1(zt^{-1};p) \,\vartheta_0(zt     ;p) \over 
  \vartheta_1(zt     ;p) \,\vartheta_0(zt^{-1};p) },\\
C 
&\!\!\!=\!\!\!&
{(1-q)(1-t^{-1})\over (1-p)f(p)}=
 {( q;p^2)_\infty (t^{-1};p^2)_\infty \over 
  (pq;p^2)_\infty (pt^{-1};p^2)_\infty},
\ea
for $|p|<1$.
Here, 
$\vartheta_1(z;p) = 
-ip^{1\over 4} z^{1\over 2}
(p^2 z;p^2)_\infty$$( z^{-1};p^2)_\infty$$(p^2;p^2)_\infty$ and
$\vartheta_0(z;p) = 
(p   z;p^2)_\infty \times$ $\times(pz^{-1};p^2)_\infty$$(p^2;p^2)_\infty$
with $(z;q)_\infty \equiv \prod_{n\geq 0}(1-zq^n)$.

Next, let $p = e^{\tau\pi i}$ 
and perform a modular transformation 
$\tau\rightarrow -1/\tau$  for theta functions
and take a limit $-1/\tau \rightarrow i\infty$, {\it i.e.}
$p\rightarrow 1$. Changing the parameterization as 
$z=p^{i{\theta\over\pi}}$ and $t=p^{\xi}$, we have 

\proclaim Proposition 2. 
\hspace{-2mm}\cite{rL}\footnote{The notations in \cite{rL} are
$x=p^{1\over 2}$, $\xi=\beta/(1-\beta)$ and $\epsilon=i\tau$.}~
In the limit of $p\rightarrow 1$,  
the $q$-Virasoro generator satisfies 
the following Zamolodchikov-Faddeev algebra\footnote{
This ZF-equation should be understood in the sense of analytic continuation.}
 of the sine-Gordon model,
\be\label{eZFSG}
{\cal T}(\theta_1) {\cal T}(\theta_2) = S\left( {\theta_1- \theta_2}\right) 
{\cal T}(\theta_2) {\cal T}(\theta_1), \qquad
S(\theta) = {\sinh\theta + i\sin\pi\xi \over \sinh\theta - i\sin\pi\xi},
\ee
where ${\cal T}(\theta)=\lim_{-1/\tau \rightarrow i\infty}T(z)$.

\noindent
Namely,
the $p=1$ $q$-Virasoro generator ${\cal T}(z)$ creates 
the basic particle (the first breather) with a rapidity $\theta$
of the sine-Gordon model, defined by the following Lagrangian density
\be\label{eLagrangianSG}
{\cal L}_{SG} = 
{1\over2}(\partial_\mu\phi)^2 + 
\left( {m_0\over b}\right)^2 \cos\left( b\phi\right),\qquad 
b^2=8\pi\beta=8\pi{\xi\over 1+\xi},
\ee
at the attractive range $0<\xi<1$.
This identification is based on the 
coincidence of the $S$-matrix for the 
basic scalar particles in sine-Gordon model with
$f(z)/f(1/z)$ in the limit of $p\rightarrow 1$, and 
the existence of
simple poles at the points $\theta_1 = \theta_2 \pm i\pi$
in the delta function terms. 
For the axiom of the ZF operator, see \cite{rL2}.
Lukyanov also proposed in \cite{rL} that,
for generic $p$,
eq.\ \eq{eq:ZFXYZ} can be  interpreted as the ZF relation 
of the XYZ model,
{\it i.e.},
the $q$-Virasoro generator is 
the basic scalar creation operator of this model.

\subsubsection{limit of $q\rightarrow 0$}
Next, study the $q\rightarrow 0$ limit ($t$: fixed).
This limit is interesting from the point of view of the 
representation theory of the Hall-Littlewood polynomials \cite{rM}
in terms of the Heisenberg algebra.
It is known by the work of Jing \cite{rJ}
that the Hall-Littlewood polynomials are 
realized by a multiple integral formula.
The integration kernel is 
simply given by multiplying vertex 
operators on a vacuum state.
As for the number of the integration variables
and the vertex operators, it is related with the 
shape of the Young diagram of each Hall-Littlewood polynomial.
His realization can be regarded as a deformation of 
the determinant representation of the Schur polynomials.
Recently, similar integral 
representations are studied
for the Jack polynomials and the 
Macdonald polynomials \cite{rSt,rM,rMY,rAMOS,rAOS}. 
However, the number of the integrals 
there is much greater in general 
than the case of the Schur or Hall-Littlewood
polynomials. We will discuss this in 
Section \ref{q0limit}
by using the Heisenberg realization of \v in the limit $q\rightarrow0$.

So as to obtain well behaving generators at $q \to 0$, 
let us scale $T_n$ as  
\be
{\tilde T}_n = T_n p^{\frac{|n|}{2}}.
 \label{r1}
\ee
Using this notation and 
taking the limit ($q \to 0$) of the relation (\ref{e:a1.2}),
we have the commutation relation for 
the deformed Virasoro algebra in this limit.
\proclaim Proposition 3. The commutation relations for 
the deformed Virasoro generators
${\tilde T }_n$ are 
\ba
\left[{\tilde T }_n,{\tilde T }_m\right] &=&
-(1-t^{-1})\sum_{\ell =1}^{n-m}{\tilde T }_{n-\ell}{\tilde T }_{m+\ell}
\quad \mbox{for}
\quad n > m > 0 \quad  \mbox{or}\quad  0>n>m,  
\n 
\left[{\tilde T }_n,{\tilde T }_0\right] &=&
 -(1-t^{-1})\sum_{\ell =1}^{n}{\tilde T }_{n-\ell}{\tilde T }_{\ell}
-(t-t^{-1})\sum_{\ell =1}^{\infty} t^{-\ell}
{\tilde T }_{-\ell}{\tilde T }_{n+\ell} \quad \mbox{for}\quad   n > 0,  
\n 
\left[{\tilde T }_0,{\tilde T }_m\right] &=&
 -(1-t^{-1})\sum_{\ell =1}^{-m}{\tilde T }_{-\ell}{\tilde T }_{m+\ell}
-(t-t^{-1})\sum_{\ell =1}^{\infty} t^{-\ell}
{\tilde T }_{m-\ell}{\tilde T }_{\ell} \quad \mbox{for}\quad   0 > m,  
\n 
\left[{\tilde T }_n,{\tilde T }_m\right] &\!\!=\!\!&
-(1-t^{-1}){\tilde T }_{m}{\tilde T }_{n} 
-(t-t^{-1})\sum_{\ell =1}^{\infty} t^{-\ell}
{\tilde T }_{m-\ell}{\tilde T }_{n+\ell} 
\n 
& & +(1-t^{-1})\mbox{\rm sign}(n)\delta_{n+m,0}
\quad \mbox{for}\quad    n> 0> m, 
\label{r2}
\ea
where the function $\mbox{\rm sign}(n)$ is 
$1$, $0$ and $-1$ for $n>0$, $n=0$ and $n<0$, respectively.


\noindent

\subsection{highest weight modules of  \v}
Let us define the Verma module of \v.
Let  $\kv{\lambda}$ be the highest weight vector
such that 
$T_0 \kv{\lambda}=\lambda\kv{\lambda}$, $\lambda\in{\bf C}$ and 
$T_n \kv{\lambda}=0$ for $n>0$. 
The Verma module $M(\lambda)$ is defined by
$M(\lambda)=\mbox{\v}\kv{\lambda}$.
The irreducible highest module $V(\lambda)$ is obtained from
$M(\lambda)$ by removing all singular vectors and their descendants.
Right modules are defined in a similar way from the
lowest weight vector
$\bv{\lambda}$
s.t.
$\bv{\lambda}T_0=\lambda\bv{\lambda}$ and
$\bv{\lambda}T_n=0$ for $n<0$. 
A unique invariant paring is defined by setting
$\bv{\lambda}\lambda \rangle = 1$.
The Verma module $M(\lambda)$ may have
singular vectors same as that of the ordinary Virasoro algebra.
Let us introduce the (outer) grading  operator $d$ which satisfies
$[d , T_n]= n T_n$ and set $d\ket{\lambda}=0$. We call a vector
$\ket{v} \in M(\lambda)$ of level $n$ if $d\ket{v}=-n\ket{v}$.

Whether there exist the singular vectors or not is
checked by calculating the Kac determinant.
Here, we give some explicit forms of $f_{n}$
which we will use for the calculations
\bea
& & f_{1}=\frac{(1-q)(1-t^{-1})}{1+p},\nonumber
\label{e:a11.2}
\\
& &
f_{2}=\frac{(1-q^{2})(1-t^{-2})}{2(1+p^{2})}+
\frac{(1-q)^2(1-t^{-1})^2}{2(1+p)^2}. \nonumber
\label{e:a11.3}
\eea
\par
At level 1, the Kac determinant is the $1\times 1$
matrix as follows
\be
\bv{\lambda}T_{1} T_{-1}\kv{\lambda}
=\frac{(1-q)(1-t)}{q + t}(\lambda^2 - (p^{1/2}+p^{-1/2})^2 ).
\label{e:a15}
\ee
Therefore, there exist a singular vector at level 1 iff
$\lambda=\pm \left(p^{1/2}+p^{-1/2}\right)$,
 since  $q$ and $t$ are generic.
The signs $\pm$ in the RHS are due to the symmetry \eq{e:a1.3}.
\par
At level 2, the Kac determinant is
\bea
\& \hskip-10truemm
 \left|
         \begin{array}{clcr}
              \bv{\lambda}T_{1}T_{1}T_{-1}T_{-1}\kv{\lambda} &
                 \bv{\lambda}T_{1}T_{1}T_{-2}\kv{\lambda} \\
              \bv{\lambda}T_{2}T_{-1}T_{-1}\kv{\lambda} &
                   \bv{\lambda}T_{2}T_{-2}\kv{\lambda} \\
         \end{array}
 \right|
=
 \frac{(1-q^2)(1-q)^2 q^{-4}(1-t^2)(1-t)^2 t ^{-4}}{(q +t)^2 (q^2 +t^2)}
\nonumber \\
\&\hskip35truemm 
\times (\lambda^2 qt-(q+t)^2 )( \lambda^2 q^2t-(q^2+t)^2 )
           ( \lambda^2 qt^2 - (q+t^2)^2 ).
\label{e:a18}
\eea
The vanishing conditions of the Kac determinant are
\bea
& \mbox{(i)}&\lambda =\pm \left(p^{1/2} + p^{-1/2}\right),
\label{e:a19.1}  \\
& \mbox{(ii)}&\lambda  =\pm \left(p^{1/2}q^{1/2} + p^{-1/2}q^{-1/2}\right),
\label{e:a19.2}  \\
& \mbox{(iii)}& \lambda =\pm \left(p^{1/2}t^{-1/2} + p^{-1/2}t^{1/2}\right).
\label{e:a19.3}
\eea
In the case (i), there is a singular vector at level 1.
In the cases (ii) and (iii), we have a singular  vector at level 2.
The singular vector for the case (ii) is
\be
 \frac{qt^{-1/2}(q+t)}{(1-q)^2(1+q)}T_{-1}T_{-1}\kv{\lambda}
\mp T_{-2}\kv{\lambda},
\label{e:a20.1}
\ee
and for (iii) is
\be
 \frac{q^{-1/2}t(q+t)}{(1-t)^2(1+t)}T_{-1}T_{-1}\kv{\lambda}
\mp  T_{-2}\kv{\lambda}.
\label{e:a21.1}
\ee

To calculate the Kac determinant becomes difficult task
when $N$ increases.
We have calculated up to level 4, and write down the
conjectural form at level $N$.

\proclaim Conjecture 1. The Kac determinant at level-N is written as 
\begin{equation}\label{Kacconj}
  \det{}_N
  =
  \det\Bigl(\langle i\ket{j}\Bigr)_{1\leq i,j\leq p(N)}
  \!\!=\!\!
  \prod_{\scriptstyle r,s\geq 1 \atop \scriptstyle rs\leq N}
  \Bigl(\lambda^2-\lambda_{r,s}^2\Bigr)^{p(N-rs)}
  \left(\frac{(1-q^r)(1-t^r)}{q^r+t^r}
  \right)^{p(N-rs)},
\end{equation}
where 
$
\lambda_{r,s}=
t^{r/2}q^{-s/2}+t^{-r/2}q^{+s/2}
$
and
the basis at level $N$ is defined
$\ket{1}=T_{-N}\ket{\lambda}$,
$\ket{2}=T_{-N+1}T_{-1}\ket{\lambda}$,$\cdots,
\ket{p(N)}=T_{-1}^N\ket{\lambda}$, and $p(N)$ is the
number of the partition of $N$.

\noindent
We remark that the $\lambda$ dependence has  essentially
the same structure as the case of the usual Virasoro algebra.
Therefore, if $q$ and $t$ are generic,  the character of the quantum Virasoro
algebra \v,
which counts the degeneracy at each level,
exactly coincides with that of the usual Virasoro algebra.
The $\lambda$-independent factor
in the RHS will play an important role when we  study the case
that $q$ is a root of unity.

\subsection{problem of obtaining a geometric interpretation of \v}
It is remarkable that \v arises in a variety of off-critical models
in a universal way.
As for the geometric interpretation, however, 
we have not 
have a satisfactory answer yet.
For the ordinary Virasoro algebra with $c=0$, we have 
the differential operator realization,
$L_n=-z^{n+1} \partial_z$, 
which explains that 
the Virasoro algebra describes the Lie algebra structure 
of the tangent space of the conformal group.
As a natural deformation of this differential operator realization,
is it possible to have a difference operator representation of
the deformed Virasoro algebra  
\v for some parameters $q$ and $p$?
It may be possible to study a connection between \v
and the analysis over the local fields in the limit of 
$q\rightarrow 0$ with fixed $t$. See Section \ref{q0limit}.

%
%
\section{Free boson realization of \v }
In this section, we present the Heisenberg realization of \v
and its applications to Calogero-Sutherland-type models.

\subsection{conformal field theory}
One of the simplest example of the conformal field theory is 
the massless Klein-Gordon field in $1+1$ dimensions. 
We briefly review how we can treat the 
Virasoro current in terms of the Klein-Gordon 
field at the conformal point to 
prepare basic ideas and notations for the later discussions.
As for the detail, the readers are referred to the original or
review articles of the conformal field theory \cite{rBPZ,rG}.
The 
action for the Klein-Gordon field $\phi(x,\tau)$ is 
$$
S_{Eucl}=\int d\tau dx{1 \over 2} \left(
 (\partial_\tau \phi)^2+ (\partial_x \phi)^2
+ m^2 \phi^2\right).
$$
If the system is massless $m=0$ then it acquires the infinite dimensional 
conformal symmetry. 
We shall see how the generators of this conformal 
transformation are realized in terms of the 
Klein-Gordon field $\phi(x,\tau)$.
Looking at the 
equation of motion
\begin{eqnarray}
\partial_w \partial_{\bar w}\phi(w,\bar{w})=0,\qquad\qquad (w=x+i \tau),
\end{eqnarray} 
we have the decoupling of $\phi$ into 
chiral and anti-chiral parts as
$$
\phi(w,\bar{w})=a(w)+\bar{a}(\bar{w}).
$$
Therefore,
we can study the chiral part and anti-chiral part separately.
After the compactification of the 
space into the segment 
$0\leq x \leq 2\pi$ with the periodic boundary condition 
and introducing the conformal mapping $z=e^{iw}$, we arrive at
the expansion
\begin{eqnarray}
a(z)=Q+a_0\ln z  - \sum_{n\neq 0} {a_n \over n} z^{-n}.
\end{eqnarray}
The Poisson brackets for these modes are
\begin{eqnarray}
&& \{ a_n,a_m\}_{\rm P.B.}=n \delta_{n+m,0}\qquad
\{ a_n,Q\}_{\rm P.B.}= \delta_{n,0}.
\end{eqnarray}
The Virasoro current $L(z)$ is written as
$$
L(z)= \sum_{n\in{\bf Z}} L_n z^{-n-2}=
{1 \over 2} \left( \partial a(z) \right)^2.
$$
Using the formula
$$
\{ \partial a(z),\partial a(w)\}_{\rm P.B.}= {1 \over z^2}\delta'(w/z),
$$
we obtain
\begin{equation}
\{ L_n,L_m\}_{\rm P.B.}= (n-m) L_{n+m}.
\end{equation}

To quantize the system, we replace the Poisson brackets by
the commutators as
\begin{equation}
[a_n,a_m]= n \delta_{n+m}\qquad [ a_n,Q]= \delta_{n,0},
\end{equation}
and define the Virasoro current with the
normal ordered product as
$$
L(z)= {1 \over 2} :\left( \partial a(z) \right)^2:.
$$
The definition of this normal ordering is that 
we shift every 
annihilation operators ({\it i.e.}, $a_n$ with $n\geq 0$)
to the right of 
creation operators ({\it i.e.}, $a_n$ with $n<0$ and $Q$).
For example $:a_{-1}a_{2}:=a_{-1}a_{2}$,
$:a_{3}a_{-2}:=a_{-2}a_{3}$ and so on.
This quantized Virasoro current obeys eq.\ \eq{e:a8} with $ c=1$.
In general the central charge $c$ 
depends on the model; $c=1/2$ for real fermion, $c=3k/(k+2)$ for 
$\widehat{su}(2)_k$
Kac-Moody algebra, for example.
One more important construction of the 
Virasoro algebra which is relevant to our later discussion is
the Feigin-Fuchs construction
$$
L(z)= {1 \over 2} :\left( \partial a(z) \right)^2:+
\alpha_0 \partial^2 a(z).
$$
For this realization we have the central charge less than one
$$
c=1-12 \alpha_0^2,
$$
if $\alpha_0$ is real.
For the later discussion we will parameterize $\alpha_0$ as
$$
\alpha_0 =\frac{1}{\sqrt{2}}\left({\sqrt{\beta}-\sqrt{1/\beta}}\right).
$$
We will see that this parameter $\beta $ has the meaning of the coupling
constant of the Calogero-Sutherland model. See Section \ref{secCS}.

\subsection{singular vectors of the Virasoro algebra}
What is very special in the representation space is
the singular vectors, which correspond to decoupled states from 
the physical space. We review some of the 
explicit formulas of the singular vectors.

The highest weight state $|h\rangle$ is defined by
$L_n |h\rangle=0$ for $n>0$ and
$L_0 |h\rangle=h |h\rangle$,
and the Verma module $M(h)$ is spanned over the highest weight state
$|h\rangle$ 
as $M(h)=\langle L_{-1},L_{-2},\cdots\rangle |h\rangle$.
The singular vector $|\chi\rangle \in M(h)$ at level $n$ is defined by
$L_n |\chi\rangle = 0$ for $n>0$ and 
$L_0 |\chi\rangle = (h+n)|\chi\rangle$.
By a standard argument, it has null norm with any states
in the Verma module;
$\langle * | \chi \rangle =0$.
The existence of such state depends crucially on the
choice of parameter $c$ and $h$.
Celebrated Kac formula shows that if they are
explicitly parameterized as eq.\ \eq{e:a9} and 
\be\label{eq:Kac}
h_{rs}=\frac{(\beta r-s)^2-(\beta-1)^2}{4\beta},
\label{eq:KacFormula}
\ee
for an arbitrary parameter $\beta(\neq0)\in\bC$
and integers $r$ and $s$ with $rs>0$, 
there exists unique (up to normalizatin) null state of level $rs$.
Some of the lower lying states can be explicitly
obtained by solving the defining conditions. 
Let $|\chi_{rs}\rangle\in M(h_{rs})$ be 
the null state at level $rs$.
For example,
we obtain,
\ba\label{eq:Lower}
|\chi_{11}\rangle&=& L_{-1}|h_{11}\rangle,\n
|\chi_{12}\rangle&=& (L_{-2}-{\beta}L_{-1}^2)|h_{12}\rangle,\\
|\chi_{22}\rangle &=& \left(L_{-4}+
\frac{2(\beta^2-3\beta+1)}{3(\beta-1)^2}L_{-3}L_{-1}-
\frac{(\beta+1)^2}{3\beta}L_{-2}^2\right.\n
&&\left.+
\frac{2(\beta^2+1)}{3(\beta-1)^2}L_{-2}L_{-1}^2-
\frac{\beta}{3(\beta-1)^2}L_{-1}^4\right)
|h_{22}\rangle, \nonumber 
\ea
and so on.

Since we have the Feigin-Fuchs realization of $L(z)$,
the singular vectors are also written in 
the bosonic creation oscillators $a_{-n}$ acting on the 
vacuum state $|\alpha \rangle$ 
({\it i.e.}, $a_n|\alpha\rangle=0$ for $n> 0$ 
and $a_0|\alpha\rangle=\alpha|\alpha\rangle$).
It is quite remarkable that all these singular vectors
can be regarded as the ``Jack symmetric polynomials''
when we replace $a_{-n}$ by the power sum $\sum_{i=1}^N x_i^n$.
As for the proof of this correspondence 
the reader is referred to \cite{rMY}.

In the Feigin-Fuchs construction, the Virasoro generators are 
\be\label{eq:coulomb}
L_n=\frac{1}{2}\sum_{m\in {\bf Z}} :a_{n+m} a_{-m}: -\alpha_0
 (n+1) a_n.
\ee
The highest weight state $|h_{rs}\rangle$ is realized as 
$\ket{\alpha_{rs}}\equiv e^{\alpha_{rs}Q}|0\rangle$,
with
\be
\alpha_{rs}=\frac{1}{\sqrt 2}\left((1+r)\sqrt \beta
-(1+s)\sqrt{1/\beta}\right).
\label{eq:alphaRS}
\ee
In terms of this free boson oscillators,
the null states (\ref{eq:Lower}) are written as,
\ba
|\chi_{11}\rangle & =
& a_{-1} |\alpha_{11}\rangle, \n
|\chi_{12}\rangle & =
& \left(a_{-2}
        + \sqrt{2\beta}a_{-1}^2\right)|\alpha_{12}\rangle, \\
|\chi_{22}\rangle & =
& \left( a_{-4}
        +\frac{4\sqrt{2\beta}}{1-\beta}
         a_{-3}a_{-1}
         -2\frac{1+\beta+\beta^2}{\sqrt{2\beta}(1-\beta)}a_{-2}^2
    -4 a_{-2}a_{-1}^2
        -\frac{2\sqrt{2\beta}}{1-\beta} a_{-1}^4
        \right)|\alpha_{22}\rangle.\nonumber
\ea

To translate these expressions into the Jack symmetric functions,
one can apply the rule,
\be
a_{-n}\rightarrow \sqrt{\frac{\beta}{2}}\sum_{i=1}^N x_i^n,
\qquad
\ket{\alpha_{rs}}\rightarrow 1.
\ee
Using this rule, we have the correspondence \cite{rMY}
\begin{equation}
|\chi_{rs}\rangle \sim J_{\{ s^r\}}(x;\beta),
\end{equation}
where the R.H.S. is the Jack symmetric polynomial for the 
rectangular diagram ${\{ s^r\}}$.

It was shown in \cite{rAMOS} that the Jack polynomials for
arbitrary Young diagrams are realized as 
the singular vector of the $W_N$ algebra.

\subsection{Calogero-Sutherland Hamiltonian, the Jack 
polynomials and the Virasoro algebra}
\label{secCS}
In this section, we explain the ``Calogero-Sutherland-Virasoro'' 
correspondence. As for the details, the readers are referred to 
\cite{rAMOS} and references therein. 

The Jack symmetric polynomials arise in 
the Calogero-Sutherland (CS) model \cite{rCS} as the wave functions of 
the excited states of this model.
After a suitable coordinate transformation, 
the Hamiltonian and momentum of this system become
\be
  {\cal H}=
  \sum_{i=1}^N D_i^2
  +\beta\sum_{i<j}\frac{x_i+x_j}{x_i-x_j}(D_i-D_j),
\qquad\quad
  {\cal P} =\sum_{i=1}^ND_i.
\label{e5}
\ee
where $D_i\equiv x_i\partial_{x_i}$. 
Their eigenfunctions are called as
the Jack symmetric polynomials $J_\lambda(x;\beta)$ 
and the eigenvalues are
\be
\epsilon_{\lambda}=
  \sum_{i=1}^M\Bigl(\lambda_i^2+\beta(N+1-2i)\lambda_i\Bigr),
\ee
with a Young diagram 
$\lambda=(\lambda_1,\cdots,\lambda_M)$,
$\lambda_i\geq\lambda_{i+1}\in\bZ_{\geq0}$.


When we consider the limit of $N\rightarrow \infty$,
$\tpsi_\lambda(x;\beta)$ must be regarded as 
a ``symmetric function'' \cite{rM}. 
It is the advantage of this limit that  
we are able to have 
the Heisenberg realization of 
the symmetric function $J_\lambda(x;\beta)$.
In other words,
the Jack symmetric functions are realized in the Fock space 
of the bosonic field \cite{rMY,rAMOS}.

Let us modify the normalization of the bosonic oscillators as
\be
  \Bigl[a_n,a_{m}\Bigr]=\frac{1}{\beta}n\delta_{n+m,0},
  \qquad\qquad [a_n,Q]={1\over \beta}\delta_{n,0},
\label{e7}
\ee
to make the ``polynomial-boson'' correspondence
simple.
Thus, the modification is ($n>0$)
\begin{eqnarray}
a_n=\sqrt{1 \over 2\beta}a^{\rm old}_n ,
\qquad a_{-n}=\sqrt{2 \over \beta}a^{\rm old}_{-n}, \qquad 
a_0=\sqrt{1 \over 2\beta}a^{\rm old}_0,\qquad
Q= \sqrt{2 \over \beta}Q^{\rm old}.
\end{eqnarray}
Correspondingly we change the notation for 
$\alpha_{rs}$ as
$$
\alpha_{rs}=\frac{1}{ 2}\left((1+r)\beta
-(1+s)\right),
$$
and write 
$|h_{rs}\rangle= \ket{\alpha_{rs}}\equiv e^{\alpha_{rs}Q}|0\rangle$
as before.

One may derive bosonized Hamiltonian 
and momentum $\widehat{\cal H}$ and $\widehat{\cal P}$
which satisfy
$
{\cal O} \pi_N \langle 0|\exp(\beta \sum_n{ a_n \over n}p_n)
=\pi_N \langle 0|\exp(\beta \sum_n {a_n \over n}p_n)\widehat{\cal O},
\label{e12}
$
where ${\cal O}={\cal H},{\cal P}$, and 
$\pi_N$ denotes the projection to the $N$-particle space.
They are given by,
\be
 \widehat{\cal H}= \beta \sum_{n>0} 
a_{-n} \,L_n + (\beta-1+\beta N-2 a_0)\, \widehat{\cal P},
\qquad\quad
\widehat{\cal P}=\beta\sum_{n=1}^\infty a_{-n}a_n.
\ee
Here $L_n$'s are the annihilation operators of
the Feigin-Fuchs construction of the Virasoro algebra with
the center $c$ in \eq{e:a9}.
Using these formulas, it is easily shown that
the singular vector $|\chi_{rs}\rangle$ of the Virasoro algebra is
proportional to 
the Jack polynomial for the Young diagram
$\lambda=\{s^r\}$ , since we have
\be
 \widehat{\cal H}|\chi_{rs}\rangle =
\epsilon_{\{s^r\}}|\chi_{rs}\rangle.
\ee
with $\epsilon_{\{s^r\}}=(\beta(N-r)+s)\,rs$.

\subsection{screening currents}
In the Feigin-Fuchs construction we are able to have 
two weight-one primary fields $S_\pm(z)$, which are called screening currents.
The condition of being weight-one primary gives us 
the equation
\begin{eqnarray}
[L_n,S_\pm(z)]= \partial_z \left( z^{n+1} S_\pm(z) \right). 
\end{eqnarray}
We have the solutions of this equation as follows
\begin{eqnarray}
  S_+(z)
  &\!\!=\!\!&
  \exp\left\{\ps{n}\beta \frac{a_{-n}}{n}z^{n}\right\}
  \exp\left\{-\ps{n}2\beta \frac{a_{n}}{n}z^{-n}
  \right\} e^{\beta Q}z^{2\beta a_0}, \label{Vscr1}\\
  S_-(z)
  &\!\!=\!\!&
  \exp\left\{-\ps{n}\frac{a_{-n}}{n}z^{n}\right\}
  \exp\left\{\ps{n}2\frac{a_{n}}{n}z^{-n}\right\}
  e^{- Q}z^{-2 a_0}. \label{Vscr2}
\end{eqnarray}

In the Fock module with the highest weight state
$\ket{\alpha_{r,s}}$, we have a singular vector $\ket{\chi_{r,s}}$
at level $rs$. By using a screening current $S_+(z)$,
$\ket{\chi_{r,s}}$ is given as follows \cite{rKM,rTK}:
\ba
  \ket{\chi_{r,s}}
  &\!\!=\!\!&
  \oint\prod_{j=1}^r\frac{dz_j}{2\pi i}\cdot
  \prod_{i=1}^r S_+(z_i)
  \ket{\alpha_{-r,s}} 
\label{eq:VirSing}\\
  &\!\!=\!\!&
  \oint\prod_{j=1}^r\frac{dz_j}{2\pi iz_j}\cdot
  \prod_{i,j=1 \atop i<j}^r(z_i-z_j)^{2\beta}\cdot
  \prod_{i=1}^rz_i^{(1-r)\beta-s}
  :\prod_{j=1}^r S_+(z_i):
  \ket{\alpha_{r,s}},
\ea
where the integration contour is the Felder's one \cite{rFe}. 
Note that there is a similar formula for representing $\ket{\chi_{r,s}}$
by using $S_-(z)$.

This method of writing the singular vectors in terms of the screening 
currents
thus gives us a systematic way to
represent the Jack polynomials for rectangular diagrams.

\subsection{Macdonald symmetric polynomials and \v}
So far, we have explained the relations among the 
Calogero-Sutherland Hamiltonian, Jack polynomials and 
the Virasoro algebra.
We want to study what will happen if we replace 
the Jack symmetric polynomial with the Macdonald symmetric
polynomial. 
It will be  shown that some of the Macdonald symmetric polynomials 
$P_\lambda(x;q,t)$ are 
related with the singular vectors of the 
quantum deformed Virasoro algebra \v 
with $p=qt^{-1}$.

The Macdonald symmetric polynomial $P_\lambda(x;q,t)$ 
is the eigenfunction of the 
Macdonald shift operator \cite{rM}
\be
  D_{q,t}=\sum_{i=1}^N \prod_{j\neq i}\frac{t x_i-x_j}{x_i-x_j} T_{q,x_i},
\label{eq:MacOp}
\ee
where $T_{q,x_i}$ is the $q$-shift operator,
\be
T_{q,x_i}f(x_1, \cdots, x_N) = f(x_1,\cdots,qx_i,\cdots,x_N).
\ee
This shift operator 
plays the same role as the CS Hamiltonian ${\cal{H}}$.
Here, a new complex deformation parameter $q$ is introduced and
$t$ is related to $\beta$ by $t=q^{\beta}$.
In the limit of $q\rightarrow 1$ with fixed $\beta$, 
the Macdonald polynomials 
reduce to the Jack polynomials.

Amazingly, one may find a closed form of the bosonized
Macdonald operator $\widehat{D}_{q,t}$ as 
\ba
  &\!\!\!\!& D_{q,t}\pi_N \langle 0|
\exp\left\{\sum_{n=1}^\infty \frac{1-t^n}{1-q^n}\frac{a_n}{n} p_n\right\}
  = \pi_N \langle 0|
\exp\left\{\sum_{n=1}^\infty \frac{1-t^n}{1-q^n}\frac{a_n}{n} p_n\right\}
   \widehat{D}_{q,t},
\label{eq:BosonMacOp}\\
  &\!\!\!\!& \widehat{D}_{q,t}= \frac{t^{N}}{t-1}
    \oint \frac{dz}{2\pi i} \frac{1}{z}
\exp{\left\{ \sum_{n=1}^\infty \frac{1-t^{-n}}{n} a_{-n} z^n   \right\}}
\exp{\left\{-\sum_{n=1}^\infty \frac{1-t^{ n}}{n}  a_{n} z^{-n}\right\}}
-\frac{1}{t-1},
\label{eMcCF}
\ea
where the commutation relations for the bosonic oscillators are deformed as
\begin{equation}
[a_n,a_m]= n \frac{1-q^{|n|}}{1-t^{|n|}}\delta_{n+m,0}.
\end{equation}

We can write the ``natural criterion'' for
the quantum deformed Virasoro algebra \v associated with 
the Heisenberg realization of the Macdonald symmetric polynomials
schematically in the following diagram.
\begin{eqnarray*}
\begin{array}{ccc}
{\rm \mbox{Calogero-Sutherland}\;Hamiltonian}  &
{\displaystyle\mathop{ \llla}^{\rm singular\;vectors}}&
L_n\;{\rm generates\; the\; Virasoro\;algebra} \\
\widehat{\cal H}=\ps{n}a_{-n}L_n+\cdots  & &\\
 L_n J_{\{s^r\}}(x;\beta)=0\quad {\rm for}\;n>0&& \\
&&\\
q\mbox{-deformation} \Biggl\downarrow &&
q\mbox{-deformation} \Biggl\downarrow \\
 && \\
 {\rm Macdonald\;difference\;operator}   &
{\displaystyle\mathop{ \llla}^{\rm singular\;vectors}}& 
T_n\;{\rm generates\; the\;} q\mbox{-Virasoro algebra \v } \\
\widehat{D}_{q,t} = \ps{n} \psi_{-n}T_{n}+\cdots &&\\
 T_n P_{\{s^r\}}(x;q,t)=0\quad {\rm for}\;n>0&& \\
 &&
\end{array}
\end{eqnarray*}
Here, $\psi_{-n}$ should be a suitable combination of 
the creation operators $a_{-m}$'s with degree $n$.

The problem is ``to make this diagram commutative.''
However, we have many unknown operators 
$\psi_{-n}$ and $T_n$. 
To have enough data to solve this problem, 
some knowledge of the $q$-deformed screening operators
must be needed.

By studying the action of the bosonized Macdonald operator
$\widehat{D}_{q,t}$,
we are able to obtain bosonized realization for some of 
the Macdonald polynomials.
Let,
\be
  \exp\left\{ \sum_{n=1}^\infty \frac{1-q^{\gamma n}}{1-q^n}
          \frac{a_{-n} }{n} z^n\right\} 
  =
  \sum_{n=0}^{\infty} \widehat Q_n^{(\gamma)} z^n  \label{macQ}
\ee
the states $\widehat Q_n^{(\gamma)}|0\rangle$ with $\gamma=\beta$ or $-1$
are the Macdonald polynomials $Q_\lambda(x;q,t)$
corresponding to the Young diagram
with single row $(n)$ or single column $(1^n)$, respectively.
As for the difference between $P_\lambda(x;q,t)$ and 
$Q_\lambda(x;q,t)$, see \cite{rM}.
We obtained other examples;
for the Young diagram with
two rows $\lambda=(\lambda_1,\lambda_2)$ or
two columns ${}^t\lambda=(\lambda_1,\lambda_2)$ we have
\ba
  \widehat Q_{(\lambda_1,\lambda_2)}^{({\gamma})}|0\rangle
  &\!\!=\!\!&
  \sum_{\ell=0}^{\lambda_2}c^{({\gamma})}(\lambda_1-\lambda_2,\ell)
  \widehat Q_{\lambda_1+\ell}^{({\gamma})}
  \widehat Q_{\lambda_2-\ell}^{({\gamma})}
  |0\rangle, \n
  c^{({\gamma})}(\lambda,\ell)
  &\!\!=\!\!&
  \frac{1-q^{\frac{\beta}{\gamma}(\lambda+2\ell)}}
       {1-q^{\frac{\beta}{\gamma}(\lambda+\ell)}}
  \prod_{j=1}^{\ell}
   \frac{1-q^{\frac{\beta}{\gamma}(\lambda+j)}}
        {1-q^{\frac{\beta}{\gamma}j}}\cdot
  \prod_{i=1}^{\ell}
  \frac{q^{\gamma}-q^{\frac{\beta}{\gamma}(i-1)}}
       {1-q^{\gamma+\frac{\beta}{\gamma}(\lambda+i)}},
\ea
with $\gamma=\beta$ or $-1$, respectively.
 The Macdonald polynomials of single hook $(n,1^m)$ are,
\be
  \widehat Q_{(n,1^{m})}|0\rangle =
  \sum_{\ell=0}^{m} \frac{1-q^{n+\ell}t^{m-\ell}}{1-q} q^{m-\ell}\:
  \widehat Q_{n+\ell}^{(\beta)}\widehat Q_{m-\ell}^{(-1)}|0\rangle.
\ee
These explicit formulas in terms of (\ref{macQ})
strongly suggest that the screening currents
for \v should be
\bea
  S_+(z)
  &\!\!=\!\!&
  \exp\left\{\ps{n}\frac{1-t^n}{1-q^n}\frac{a_{-n}}{n}z^{n}\right\}
  \exp\left\{{\rm annihilation\; part\; for }S_+
  \right\} e^{\beta Q}z^{2\beta a_0},\\
  S_-(z)
  &\!\!=\!\!&
  \exp\left\{-\ps{n}\frac{a_{-n}}{n}z^{n}\right\}
  \exp\left\{{\rm annihilation \;part \;for } S_- \right\}
  e^{- Q}z^{-2 a_0}.
\eea
Note that this can be regarded as a good deformation of the
screening currents (\ref{Vscr1}) and (\ref{Vscr2}). 
It is assumed that 
the zero-mode parts are not deformed.
\\

In \cite{rSKAO}, it is shown that the problem of finding 
all these unknown parts for $\widehat{D}_{q,t}$, $\psi_n$ and 
$S^\pm(z)$ is ``uniquely solved'' if we start from a nice 
ansatz for the operators. The reason or mechanism
behind the process of solving the problem have not been 
well investigated yet and seem somehow mysterious. 
What is clear so far is that the ``locality'' for  
the operators seems quite important, {i.e.}, 
the behaviors of the $\delta$-functions
in the OPE factors must be well controlled in some sence.
The mathematical structure of this ``locality'' should be 
understood in the future.
Therefore,
we will not show the processes of solving this 
problem here.
Let us just summarize the results.
\proclaim Theorem 1. \hspace{-2mm}\cite{rSKAO} 
The quantum deformed Virasoro current $T(z)$ is realized as
\bea\label{eq:qVirFFR1}
T(z)&=& \Lambda^+(z)+\Lambda^-(z),
\eea
where
\bea
\Lambda^+(z)
 &\!\!\!\!\! = \!\!\!\!& p^{1/2}\exp\left\{-\ps{n}\frac{1-t^n}{1+p^n}
\frac{a_{-n}}{n}z^{n}t^{-n}p^{-n/2}\right\}
    \exp\left\{-\ps{n}(1-t^n)\frac{a_{n}}{n}z^{-n}p^{n/2}\right\}
    q^{\beta a_{0}}, \nonumber  \\
\Lambda^-(z)
  &=&\!\!\! p^{-1/2}\exp\left\{\ps{n}\frac{1-t^n}{1+p^n}
\frac{a_{-n}}{n}z^{n}t^{-n}p^{n/2}\right\}
    \exp\left\{\ps{n}(1-t^n)\frac{a_{n}}{n}z^{-n}p^{-n/2}\right\}
    q^{-\beta a_{0}}.
\label{e:b1}
\eea

\noindent By studying the operator product expansions, it is 
easy to see that
this $T(z)$ satisfies the relation of \v (\ref{e:a1.2}).
We can observe that this formula has strong resemblance to the
dressed vacuum form (DVF) in the algebraic Bethe ansatz.
This profound $q$-Virasoro-DVF correspondence 
($T(z)=\Lambda^+(z)+\Lambda^-(z)$) was
discovered by Frenkel and Reshetikhin \cite{rFR}.

\proclaim Theorem 2. \hspace{-2mm}\cite{rSKAO}  
The Macdonald operator is written as
\bea
\widehat{D}_{q,t}&=&\frac{t^N}{t-1}\left[\oint\frac{dz}{2\pi\i}\frac{1}{z}
\psi(z)T(z) -p^{-1}q^{-2\beta a_0}\right] -\frac{1}{t-1}
\nonumber \\
&= &\frac{t^N}{t-1}\left[\pzs{n}\psi_{-n}T_n
             -p^{-1}q^{-2\beta a_0}\right]-\frac{1}{t-1},
\label{e:b7.3}
\eea
where the field $\psi(z)$ is given by
\be
\psi(z)=\pzs{n}\psi_{-n}z^n=p^{-1/2}\exp\left\{-\ps{n}\frac{1-t^n}{1+p^n}
\frac{a_{-n}}{n}z^np^{n/2}t^{-n}\right\}
                      q^{-\beta a_0 }.
\ee

\proclaim Theorem 3. \hspace{-2mm}\cite{rSKAO} The screening currents for \v
\bea
  S_+(z)
  &\!\!=\!\!&
  \exp\left\{\ps{n}\frac{1-t^n}{1-q^n}\frac{a_{-n}}{n}z^{n}\right\}
  \exp\left\{-\ps{n}(1+p^n)\frac{1-t^n}{1-q^n}\frac{a_{n}}{n}z^{-n}
  \right\} e^{\beta Q}z^{2\beta a_0},
  \label{e:c1.1}\\
  S_-(z)
  &\!\!=\!\!&
  \exp\left\{-\ps{n}\frac{a_{-n}}{n}z^{n}\right\}
  \exp\left\{\ps{n}(1+p^n)\frac{a_{n}}{n}z^{-n}p^{-n}\right\}
  e^{- Q}z^{-2 a_0},
  \label{e:c1.2}
\eea
satisfies the 
commutation relation:
\bea
  \Bigl[T_n,S_+(w)\Bigr]
  &\!\!=\!\!&
  -(1-q)(1-t^{-1})\frac{d_q}{d_q w}
  \left((p^{-\frac12}w)^{n+1}p^{\frac12}:\Lambda^-(p^{-\frac12}w)S_+(w):
  \right), \label{e:c2.1}\\
  \Bigl[T_n,S_-(w)\Bigr]
  &\!\!=\!\!&
  -(1-q^{-1})(1-t)\frac{d_t}{d_t w}
  \left((p^{\frac12}w)^{n+1}p^{-\frac12}:\Lambda^+(p^{\frac12}w)S_-(w):
  \right),
  \label{e:c2.2}
\eea
where
the difference operator with one parameter
is defined by
\be
\frac{d_\xi}{d_\xi z}g(z)=\frac{g(z)-g(\xi z)}{(1-\xi)z}.
\label{e:c6}
\ee

Since the Kac determinant has the same structure as 
the ordinary Virasoro algebra has, the structure of the
singular vectors are not changed in an essential manner.
As for the case of the Jack polynomials, 
we can write down all the singular vectors at least
formally in the following way
\ba\label{eq:qVirSing}
  \ket{\chi_{r,s}}
  &\!\!=\!\!&
  \oint\prod_{j=1}^r\frac{dz_j}{2\pi i}\cdot
  \prod_{i=1}^r S_+(z_i)
  \ket{\alpha_{-r,s}} ,
\ea
however, we have to be very careful about the 
integration cycle.
Recently, very nice construction of 
the integration cycles together with
some modification of the integration kernel
is achieved in \cite{rLP2,rJLMP}.
A $q$-deformation of the Felder complex is 
constructed in these works, and 
mathematical treatment becomes much ``easier'' than
the original case.

Finally, we have as the Jack case,

\proclaim Theorem 4.
There exists a one to one correspondence between 
the singular vectors $|\chi_{r,s}\rangle $ of the $q$-Virasoro algebra \v and
the Macdonald functions $P_{\{ s^r\}}(x;q,t)$ 
with the rectangular Young diagram ${\{ s^r\}}$ up to 
normalization constants. 
It is simply given by
\begin{eqnarray}
P_{\{ s^r\}}(x;q,t) \propto 
 \langle  \alpha_{r,s}| 
\exp\left\{\sum_{n=1}^\infty \frac{1-t^n}{1-q^n}\frac{a_n}{n}
p_n\right\}|\chi_{r,s}\rangle,
\end{eqnarray}
where $\langle  \alpha_{r,s}| \alpha_{r,s}\rangle=1$.

For the general Young diagram, 
the Macdonald polynomials correspond to
the singular vectors of $q$-${\cal W}$ algebras \cite{rAKOS}.

\subsection{\v and the Hall-Littlewood polynomial}
\label{q0limit}
If we take the limit $q \to 0$, the Macdonald 
polynomial $P_\lambda(x;q,t)$ reduces to the 
Hall-Littlewood polynomial $P_\lambda(x;t)$ \cite{rM}.
Let us study the limit of \v in $q \to 0$,
and the connection 
between \v and the Hall-Littlewood polynomials $P_\lambda(x;t)$.
The commutation relations are already given in (\ref{r2}). 
The Kac determinants at lower levels are calculated as
\ba
 \det{}_1 &\!\!=\!\!&
 \langle \lambda| {\tilde T}_1 {\tilde T}_{-1}|\lambda\rangle =1-t^{-1}, \n 
 \det{}_2
  &\!\!=\!\!&
  \left|
  \begin{array}{cc}
    \langle\lambda| {\tilde T}_{2}{\tilde T}_{-2}|\lambda\rangle &
    \langle\lambda| {\tilde T}_{2}{\tilde T}_{-1}{\tilde
T}_{-1}|\lambda\rangle \\
    \langle\lambda| {\tilde T}_{1}{\tilde T}_{1}{\tilde T}_{-2}|\lambda\rangle &
    \langle\lambda| {\tilde T}_{1}{\tilde T}_{1}{\tilde T}_{-1}{\tilde
T}_{-1}|\lambda\rangle
  \end{array}
  \right| \n
  &\!\!=\!\!&
  (1-t^{-1})^2 (1-t^{-2}).
  \label{r5}
\ea
Here,
we observe that the Kac determinants do not depend on $\lambda$.
Therefore, if $t$ is generic, we have no singular vectors for any $\lambda$.
To study the degeneration of the boson realization, 
we have to restrict the zero-mode charge of the vacuum 
$|\alpha_{r,s}\rangle$ to $s= 0$, otherwise, we are not able to 
obtain nontrivial algebra.
It is easy to see that the operators
$\tilde{\Lambda}^\pm(z)\equiv
\lim_{q\rightarrow0}\Lambda^\pm(p^{\pm1/2}z)$ are well behaving ones.
Introducing the renormalized boson
$b_n=- t^{n} a_n$, $b_{-n}= a_{-n}$ for $n>0$, 
we have in $q\rightarrow0$, 
\be
[b_n,b_m]=n {1 \over 1-t^{-|n|}}\delta_{n+m,0},
\ee
\bea
\tilde{\Lambda}^+(z)
 &=&  t^{r/2} \exp\left\{\ps{n} (1-t^{-n})
\frac{b_{-n}}{n}z^{n} \right\}
    \exp\left\{-\ps{n}(1-t^{-n})\frac{b_{n}}{n}z^{-n}\right\}, 
\label{eq:BosonHall} \\
\tilde{\Lambda}^-(z)
  &=&  t^{-r/2} \exp\left\{-\ps{n}(1-t^{-n})
\frac{b_{-n}}{n}z^{n}\right\}
    \exp\left\{\ps{n}(1-t^{-n})\frac{b_{n}}{n}z^{-n}\right\},\nonumber  
\eea
on the Fock space spanned over $|\alpha_{r,0}\rangle$.
Note that these are essentially the same as 
Jing's operators $H(z)$ and $H^*(z)$ for the 
Hall-Littlewood polynomial $P_\lambda(x;t^{-1})$ \cite{rJ}.
Using this notation, the rescaled \v generator
$ {\tilde T }_n$ is expressed as
\be
{\tilde T }_n =\oint\frac{dz}{2 \pi i z}
\left(\,\, \theta[\,n\leq0\,] \tilde{\Lambda}^+(z) + 
           \theta[\,n\geq0\,] \tilde{\Lambda}^-(z)
\,\,\right) z^n,
\ee
where $\theta[P]=1$ or $0$ 
if the proposition $P$ is true or false, respectively. 
This formula and the coincidence of our
$\tilde{\Lambda}^+(z)$ with Jing's $H(z)$ means that 
in $q\rightarrow0$ limit,
``all the vectors in the Fock module'' are
written in terms of the Hall-Littlewood polynomials.

The screening currents will disappear in this limit
in the sense that $S_-(z)$ and
$\left[\tilde{T}_n,S_+(w)\right]$ become singular.
The disappearance of the singular vectors in the Fock module 
which is derived from the study of the Kac determinants
is explained by this singular behavior of the 
screening currents.

It seems interesting to study the relation between \v and the
Hall algebra \cite{rM} which is related with the analysis over 
the local fields.
Is it possible to have a geometric interpretation 
of \v for $q=0$?


\section{Further aspects of \v and relations with other models}


Here we discuss {\it i)}
the representation theories, {\it ii)} application to the ABF model and 
{\it iii)} a hidden elliptic algebra generated by screening current.
We will also study the limits of $\beta=1$, $3/2$ and $2$,
and find some connections with the Kac-Moody algebras
when $\beta=1$, for $\beta=3/2$, the $q$-Virasoro generator
is given by a BRST exact form and construct a topological model,
and when $\beta=2$, the $q$-Virasoro algebra relates
with $c=1$ ${\cal W}_{1+\infty}$ algebra. 


\subsection{symmetric realization of \v and vertex operators}
\label{secSYMMETRIC}

When we introduced the Feigin-Fuchs realization,
the creation operators and the annihilation operators
had nice symmetry. However we destroyed this symmetry 
to make many formulas for the Jack and Macdonald 
symmetric polynomials become simple.
We will study some applications of \v to other 
solvable systems. So, it helps us very much
to have a ``symmetric expressions'' of 
$T(z)$ and $S_\pm(z)$ \cite{rFFr,rAKOS,rLP2,rKa,rAKMOS}.

Let us introduce the fundamental Heisenberg algebra
$h_n$ ($n\in{\bf Z}$), $Q_{h}$ having the commutation relations
\ba
  [h_n,h_m]
  &\!\!=\!\!&
  \frac{1}{n}\frac{(q^{\frac{n}{2}}-q^{-\frac{n}{2}})
  (t^{\frac{n}{2}}-t^{-\frac{n}{2}})}{p^{\frac{n}{2}}+p^{-\frac{n}{2}}}
   \delta_{n+m,0}
   ,\qquad
  [h_n,Q_h]=\frac12\delta_{n,0}. \label{boscom}
\ea
The correspondence to the bosonic oscillators 
in the last two subsections is $(n>0)$
\be
h_{ n}={t^n -1 \over n} a_{ n},\qquad
h_{-n}= {1\over n}{1-t^{-n}\over 1+p^n} a_{-n},\qquad
h_0 = \sqrt\beta a_0,\qquad
Q_h ={\sqrt\beta\over 2} Q.
\ee
By these, 
the Virasoro current $T(z)$ and the screening current  $S_{\pm}(z)$
(with some modification) are written as
\ba
  T(z) &\!\!\!=\!\!\!& \Lambda^+(z)+\Lambda^-(z),
\label{eq:qVirFFR2}\\
  \Lambda^\pm(z) &\!\!\!=\!\!\!&
  :\exp\left\{\pm\sum_{n\neq 0}h_np^{\pm\frac{n}{2}}z^{-n}\right\}:
  q^{\pm\sqrt{\beta}h_0}p^{\pm\frac12}, \\
  S_{+}(z) &\!\!\!=\!\!\!& 
  :\exp\left\{- \sum_{n\neq 0}
  \frac{p^{\frac{n}{2}}+p^{-\frac{n}{2}}}
 {q^{\frac{n}{2}}-q^{-\frac{n}{2}}}
  h_nz^{-n}\right\}:
  e^{2\sqrt{\beta}Q_h}z^{2\sqrt{\beta}h_0},
\label{eq:SC21}\\
  S_{-}(z) &\!\!\!=\!\!\!& 
  :\exp\left\{ \sum_{n\neq 0}
  \frac{p^{\frac{n}{2}}+p^{-\frac{n}{2}}}
 {t^{\frac{n}{2}}-t^{-\frac{n}{2}}}
  h_nz^{-n}\right\}:
  e^{-{2\over\sqrt{\beta}}Q_h}z^{-{2\over\sqrt{\beta}}h_0}.
\label{eq:SC22}
\ea
If we introduce the isomorphisms $\theta$ and $\omega$ 
of the Heisenberg algebra
related with (\ref{e:aa1}), (\ref{e:aa2}):
\ba
  &&{\rm (II')}\;\;\;\; \theta:
 (q,t) \mapsto (q^{-1},t^{-1}),\quad 
  h_n \mapsto  -h_n \:(n\neq 0),
\quad
  h_0 \mapsto h_0, \quad Q_h \mapsto Q_h, \cr
  && {\rm (III')} \;\;\;\;\omega: \quad 
 q \leftrightarrow t \quad,  \quad 
  h_n \mapsto -h_n,
  \quad Q_h \mapsto - Q_h , \label{symm}
\ea
then
$\Lambda^-(z) = \theta\cdot \Lambda^+(z) =  \omega\cdot \Lambda^+(z)$,
$S_{-}(z) = \omega\cdot S_{+}(z)$ and
$ \theta\cdot S_{\pm}(z)= S_{\pm}(z)$.
Here $\omega\cdot \beta$ should be understood as $1/\beta$.
Under the isomorphism $\sigma$ such that:
\be
  {\rm (IV)}\;\;\;\; \sigma:
  q \leftrightarrow 1/t, \qquad 
  \sqrt\beta \leftrightarrow -\sqrt{1/\beta},
\ee
$\sigma\cdot\Lambda^\pm(z) = \Lambda^\pm(z)$ and
$\sigma\cdot S_\pm(z) = S_\mp(z)$.

The free boson realization for $T(z)$ is expressed as 
the following deformed Miura transformation \cite{rFR}
\be
:\!\left( p^D - \Lambda^+(z) \right) \left( p^D - \Lambda^-(z) \right)\!:\,
= p^{2D} - T(z) p^D + 1,
\label{eq:qMiura}
\ee
which has been generalized to define 
the $q$-deformed $\cal W$ algebra \cite{rFFr,rAKOS}.
By using this transformation,
Frenkel-Reshetikhin \cite{rFR} proposed a generalization of
their quasi-classical $q$-Virasoro algebra to $ABCD$-type cases.
An analogy to the Baxter's dressed vacuum form $Q$ defined by
$:\!\left( p^D - \Lambda^-(z) \right) Q(z)\!:\, = 0$,
so $\Lambda^\pm(z) = \,:\! Q(zp^{\mp1})\, Q(z)^{-1}\!:$,
seems to be of some interest.


The vertex operator defined by
\be
V_{2,1}(z) \equiv \,:\! \exp\left\{
\sum_{n\neq0}{h_n\over q^{n\over2} - q^{-{n\over2}} } z^{-n} \right\}\!:
e^{-\sqrt\beta Q_h} z^{-\sqrt\beta h_0},
\label{eq:Vertex21}
\ee
satisfies
$\theta\cdot V_{2,1}(z) = V_{2,1}(z)$ and
\ba
g\left({w\over z}p^{\pm{1\over2}}\right) T(z) V_{2,1}(w) 
&\!\!\!-\!\!\!&
V_{2,1}(w) T(z) g^{-1}\left({z\over w}p^{\mp{1\over2}}\right)\\
&\!\!\!=\!\!\!&
t^{\pm{1\over4}}(t^{1\over2} - t^{-{1\over2}})\,
\delta\left(t^{\pm{1\over2}}{w\over z}\right)
 V_{2,1}(q^{\pm{1\over2}} w) p^{\mp{1\over2}},
\label{eq:T&V21}
\ea
with
\be
g(x) = 
t^{\pm{1\over4}}
\exp\left\{\pm\sum_{n>0}{1\over n}
{ t^{n\over2} - t^{-{n\over2}} \over p^{n\over2} - p^{-{n\over2}} }x^n
\right\}.
\ee
Note that
$ V_{2,1}(q^{\pm{1\over2}} w) p^{\mp{1\over2}} =
\,:\! \Lambda^\mp(t^{\mp{1\over2}}w) V_{2,1}(w)\!:\,$. 
If we let
$V_{1,2}(z) \equiv \sigma\cdot V_{2,1}(z)$ and their fusion as
\be
V_{\ell+1,k+1}(z) \equiv \,:\!
\prod_{i=1}^\ell V_{2,1}(q^{\ell+1-2i\over2\ell}z)
\prod_{j=1}^k    V_{1,2}(t^{k   +1-2j\over2k   }z)
\!:,
\label{eq:VertexLK}
\ee
then they also obey a similar commutation relation as \eq{eq:T&V21}, 
which reduces to the usual defining relation for the Virasoro primary field
of the conformal weight $h_{\ell+1, k+1}$,
in the limit $q\rightarrow1$ \cite{rAKMOS}.
The adjont action of the \v generator $T(z)$ 
on this fused vertex operator $V_{\ell+1,k+1}$
may be closely connected with 
a coproduct of the algebra \v.
Similar but slightly different definition for fused vertex operators
\begin{equation}
  :\!
  \prod_{i=1}^{\ell}V_{2,1}(q^{-\frac{k}{2}}t^{\frac{\ell+1}{2}-i}z)
  \prod_{j=1}^k V_{1,2}(t^{-\frac{\ell}{2}}q^{\frac{k+1}{2}-j}z)
  \!:, \label{kade}
\end{equation}
was proposed in \cite{rKa}. 
The meaning of fused operators given by (\ref{eq:VertexLK}) or
(\ref{kade}) has not been made clear yet.

The fundamental vertex operators $V_{2,1}(z)$ and $V_{1,2}(z)$,
that satisfy fermion like anti-commutation relation,
are especially important.
Because
the $q$-Virasoro generator and screening currents are expressed by them
as follows;
\be
 \Lambda^+(zp^{1\over 2}) 
=\,\, : V_{2,1}^+(zq^{-{1\over 2}}) V_{2,1}^{-}(zq^{1\over 2}) : 
p^{1\over 2},\qquad
  S^+(z) =\,\, : V_{2,1}^{-}(zp^{1\over 2}) V_{2,1}^{-}(zp^{-{1\over 2}}) :,
\label{eq:TSbyV21}
\ee
and the relations obtained by $\omega$.
Here
$V^\pm_{\ell+1,k+1}(z)\equiv V^{\pm1}_{\ell+1,k+1}(z)$.
Moreover, 
the boson and power-sum correspondence operator in eq.\ \eq{eq:BosonMacOp}
is also realized as $:\prod_{i=1}^N V_{2,1}(q^{1\over2} x_i^{-1}):$.
Hence, they must play more important role in the $q$-Virasoro algebra.


\subsection{ABF model and \v}


In the papers \cite{rLP2}, the 
explicit formula for the multipoint 
correlation functions is successfully obtained. 
We review their method and the relation to \v.
%
%
%

The $q$-Virasoro algebra can be applied to the off-critical phenomena,
especially to the ABF model in the regime 
\uppercase\expandafter{\romannumeral3} \cite{rABF}.
Let $z\equiv p^v$ and the vertex operators $\Phi_\pm(z)$ be
\be
\Phi_+(z)\equiv V_{2,1}(z),\qquad
\Phi_-(z') \equiv 
 \oint{dz\over 2\pi iz} V_{2,1}(z') S_+(z) z^\beta f(v-v',\pi),
\label{e:ABFVertex}
\ee
where
$[v] \equiv p^{{1\over2} ((1-\beta)v^2-v) }
(z;q)_\infty (qz^{-1};q)_\infty (q;q)_\infty $
and 
\be
f(v,w)\equiv
{[v+{1\over2}-w]\over[v-{1\over2}]},\qquad
\pi\equiv 
-{2h_0 \over \sqrt\beta-\sqrt{1/\beta}}.
\ee
The integration contour is a closed curve around the origin satisfying 
$p|z'|<|z|<p^{-1}|z'|$.

Then the vertex operators satisfy the following commutation relation;

\proclaim Theorem 5. \hspace{-2mm}\cite{rLP2}
\footnote{The notations in \cite{rLP2} are
$x=p^{1\over 2}$ and $r=1/(1-\beta)$.}~
\be
\Phi_{\ell_3-\ell_2} (z_1)
\Phi_{\ell_2-\ell_1} (z_2)
=
\sum_{\ell_4} 
W\left(\matrix{\ell_3 & \ell_4 \cr 
               \ell_2 & \ell_1 \cr}\Bigg|\, {z_1\over z_2}\right)
\Phi_{\ell_3-\ell_4} (z_2)
\Phi_{\ell_4-\ell_1} (z_1),
\label{eq:ExchangeABF}
\ee
where $\ell_{i+1}-\ell_i = \pm 1$ with $\ell_5 \equiv \ell_1$. 
Here $W(\ell|z)$ 
is the Boltzmann weight of the ABF model such that
\ba
W\left(\matrix{\ell\pm 2 & \ell\pm 1 \cr
               \ell\pm 1 & \ell      \cr}\Bigg|\,z\right)
&\!\!\!=\!\!\!& 
R(z), \qquad\qquad\qquad
R(z)=z^{\beta\over 2} {g(z^{-1})\over g(z)},\cr
W\left(\matrix{\ell      & \ell\pm 1 \cr
               \ell\pm 1 & \ell      \cr}\Bigg|\,z\right)
&\!\!\!=\!\!\!& 
R(z) {[\ell\pm v][1] \over [\ell][1-v]}, \qquad
g(z)=
{(pz;q,p^2)_\infty (pqz;q,p^2)_\infty \over
 (qz;q,p^2)_\infty (p^2z;q,p^2)_\infty},\cr
W\left(\matrix{\ell      & \ell\mp 1 \cr
               \ell\pm 1 & \ell      \cr}\Bigg|\,z\right)
&\!\!\!=\!\!\!& 
- R(z) {[\ell\pm 1][v] \over [\ell][1-v]},
\label{eq:Boltzmann}
\ea
with $(z;q,p)_\infty \equiv \prod_{n,m\geq 0}(1-zq^np^m)$.

Besides this commutation relation 
these vertex operators also satisfy the other defining relations,
homogeneity and normalization condition \cite{rFJMMN}, 
of those of the ABF model in the regime III;
$0<p<z<1$.
Here $p=0$ and $1$ correspond to 
the zero temperature and the critical point, respectively.
Thus these vertex operators of the $q$-Virasoro algebra can be regarded as
those of the ABF model.


The exchange relation \eq{eq:ExchangeABF} was generalized 
to general vertex operators by \cite{rKa,rAKMOS}, 
and to $sl(n)$ RSOS model by \cite{rAJMP}.


\subsection{Felder resolution and the space of ABF model}


Not only the vertex operators themselves but also 
the Hilbert space on which the physical operators act 
are able to be identified with that of the ABF model.
The Hilbert space of the ABF model can be constructed through
a deformed Felder-type BRST resolution.

Let $\cF{\SSS}{\RR}$ be the Fock module 
generated by the highest weight state $|\RR,\SSS\rangle$ such that
\be
h_{n>0}|\RR,\SSS\rangle = 0,\qquad
h_0    |\RR,\SSS\rangle = 
-{1\over2}\left( \RR\sqrt\beta - \SSS\sqrt{1/\beta} \right)|\RR,\SSS\rangle.
\ee
Suppose 
$\beta = \QQ/\PP$ with coprime integers $\PP>\QQ\in\bN$ and
let the screening charge 
$\SC_+\,:\,\cF{\SSS}{\RR}\rightarrow\cF{\SSS}{\RR-2}$ be
\be
\SC_+= \oint{dz\over 2\pi iz} S_+(z) z^\beta f(v,\pi),
\ee
and define the BRST charges $Q^+_j$ ($j\in\bZ$) as
\ba
Q^+_{2j  }  =
\SC_+^\RR      \,\,\,\,&:&
\cF{\SSS}{ \RR-2j\PP}\,\,\,\rightarrow\,\,\cF{\SSS}{-\RR-2 j   \PP},\cr
Q^+_{2j+1}  =
\SC_+^{\PP-\RR}\!\!\!\!&:&
\cF{\SSS}{-\RR-2j\PP}      \rightarrow\,\,\cF{\SSS}{ \RR-2(j+1)\PP}.
\label{eq:BRST}
\ea
We also define the dual screening charge
$\SC_-\,:\,\cF{\SSS}{\RR}\rightarrow\cF{\SSS-2}{\RR}$ and
the dual BRST charges $Q^-_j$
by the replacement 
$\sqrt\beta\leftrightarrow -\sqrt{1/\beta}$, 
$q\leftrightarrow 1/t$ and
$\RR\leftrightarrow \SSS$.

\proclaim Proposition 4. \hspace{-2mm}\cite{rLP2,rJLMP}
The screening charges $\SC_\pm$ commute with each other and 
with $q$-Virasoro generators
\ba
[\SC_+,\SC_-] 
&\!\!\!=\!\!\!&
0,\cr
[T(z), \SC_\pm^{n_\pm}] 
&\!\!\!=\!\!\!&
0, \quad 
{\rm on}\,\,\,\, \cF{\RR_-}{\RR_+},\quad
{\rm with}\,\,\,\, n_\pm\equiv \RR_\pm\,\, {\rm mod} \,\,P_\pm,
\ea
and are also nilpotent
\be
Q^\pm_j Q^\pm_{j-1} 
=
\SC_\pm^{P_\pm} = 0, \quad P_\pm>1.
\label{eq:nilpotent}
\ee

Hence we can construct Felder type BRST complexes, 
for example, by $\SC_+$
\be
\cdots
\BRS{\SC_+^{    \RR}}\cF{\SSS}{-\RR+2\PP}
\BRS{\SC_+^{\PP-\RR}}\cF{\SSS}{ \RR}
\BRS{\SC_+^{    \RR}}\cF{\SSS}{-\RR}
\BRS{\SC_+^{\PP-\RR}}\cF{\SSS}{ \RR-2\PP}
\BRS{\SC_+^{    \RR}}
\cdots.
\label{eq:Felder}
\ee
{}From the Kac determinant in (\ref{Kacconj}),
the Fock module $\cF{\SSS}{\RR}$ with $\RR$, $\SSS\in\bN$ is reducible. 
To obtain an irreducible one $\cL{\SSS}{\RR}$, 
we have to factor out the submodules by the Felder resolution.
In a special case, this irreducible module coincides with 
the space of the ABF model.


To see this, we have to introduce a grading operator,
which plays the role of the corner Hamiltonian in the ABF model,
\be
H_c = \sum_{n>0} n^2 
 {p^{n\over2}+p^{-{n\over2}}\over
\left( q^{n\over2}-q^{-{n\over2}}\right) 
\left( t^{n\over2}-t^{-{n\over2}}\right) }
h_{-n} h_n + h_0^2 -{1\over 24}.
\label{eq:cornerHamil}
\ee
This commutes with screening currents up to a total divergence
\be
[H_c,S_\pm(z)] z^{\beta^{\pm1}} 
= {\partial\over\partial z} \left( S_\pm(z) z^{\beta^{\pm1}}\right),
\ee
and its eigenvalues $\EPS{\SSS}{\RR}$ 
on the Fock module $\cF{\SSS}{\RR}$ are
\be
\EPS{\SSS}{\RR}=\DEL{\SSS}{\RR} - {c\over 24} + n,\qquad 
n\in\bZ_{n\geq0},
\ee
with $\DEL{\SSS}{\RR}$ in eq.\ \eq{eq:KacFormula}.
When $\QQ=\PP-1$,
these values coincide with the eigenvalues of 
the corner Hamiltonian of the ABF model
corresponding to the $1$-$d$ configurations given by the rule:
{\it i}) each height takes an integer value between 
$1$ and $\PP-1$, {\it ii}) the allowed values of difference
in any neighboring heights are $\pm1$, {\it iii})
the height at the origin is $\RR$,
 {\it iv}) the asymptotic configuration is  
$\cdots,\SSS,\SSS+1,\SSS,\SSS+1,\cdots$.
However, we should note that the 
multiplicities of the bosonic Fock space and 
that of ABF model are different.

Lukyanov and Pugai \cite{rLP1,rLP2} showed that,
after the Felder-type BRST resolution by the dual screening current $S_-(z)$,
the multiplicities in the irreducible Fock module $\cL{\SSS}{\RR}$ 
coincide those of the ABF model.
Therefore, ABF model is completely described by 
the representation of the $q$-Virasoro algebra
and the multi-point local height probabilities of ABF model \cite{rFJMMN} are
realized as correlation functions of the vertex operators.
For example, the probability that 
the heights at the same vertical column sites have the values 
$1\leq\RR_1,\RR_2, \cdots,\RR_n\leq\PP-1$
is proportional to
\be
Tr_{\cL{\SSS}{\RR_1}}
\left[ p^{2H_c}
\Phi_{-\sigma_1    }(z_1    /p)\cdots
\Phi_{-\sigma_{n-1}}(z_{n-1}/p)
\Phi_{ \sigma_{n-1}}(z_{n-1}  )\cdots
\Phi_{ \sigma_1    }(z_1      )
\right],
\ee
where 
$\sigma_s = \RR_{s+1} - \RR_s$.


\subsection{elliptic algebra generated by the screening currents and 
$k=1$ affine Lie algebra}


The properties of screening currents are quite important 
in the representation theory of the infinite-dimensional algebra;
they govern the irreducibility and the physical states
as mentioned above subsections.
Moreover they relate with hidden quantum symmetries.


\subsubsection{an elliptic algebra generated by $S^\pm(z)$}


Here, we show that the
screening currents generate an elliptic hidden symmetry,
which reduces to the (quantum) affine Lie algebra with a special center
when $c$ tends to $1$.
Let us introduce a new current 
$\Psi(z) \equiv \,:\!S_+(q^{\pm{1\over2}}z)\, S_-(t^{\pm{1\over2}}z)\!:$, 
{\it i.e.},
\be
\Psi(z) 
= \exp\left\{ \sum_{n\neq0} 
{ p^n-p^{-n} \over (\QINT qn)(\QINT tn) } h_n z^{-n} \right\}
e^{2\alpha Q} z^{2\alpha h_0},
\ee
with $\alpha = \sqrt\beta - 1/\sqrt\beta$, 
then we have

\proclaim Proposition 5.
Screening Currents $S_\pm(z)$ and $\Psi(z)$ generate 
the following elliptic two-parameter algebra;
\ba
f_{00}\left({w\over z}\right) \Psi(z) \Psi(w) 
&\!\!\!=\!\!\!&
\Psi(w) \Psi(z) f_{00}\left({z\over w}\right),
\label{eq:ElliPP}\\
f_{0\pm}\left({w\over z}\right) \Psi(z) S_\pm(w) 
&\!\!\!=\!\!\!&
S_\pm(w) \Psi(z) f_{\pm0}\left({z\over w}\right),
\label{eq:ElliPS}\\
f_{\pm\pm}\left({w\over z}\right) S_\pm(z) S_\pm(w) 
&\!\!\!=\!\!\!&
S_\pm(w) S_\pm(z) f_{\pm\pm}\left({z\over w}\right),
\label{eq:ElliSS1}
\ea
\be
\left[\,S_+(z), S_-(w)\,\right]
= 
{1\over (p-1)w} 
\left[\, 
\delta\left(p^{  1\over2 }{w\over z}\right) \Psi(t^{-{1\over2}}w) -
\delta\left(p^{-{1\over2}}{w\over z}\right) \Psi(q^{-{1\over2}}w)
\,\right],
\label{eq:ElliSS2}
\ee
where
$f_{00}(x) =
f_{++}(x) f_{+-}^2(xp^{1\over2}) f_{--}(x)$ and 
$f_{0\pm}(x) =
f_{\pm0}(x) =
f_{+\pm}(xq^{1\over2}) f_{-\pm}(xt^{1\over2})$ 
with
\ba
f_{+-}(x) 
=
f_{-+}(x) 
&\!\!\!=\!\!\!&
\exp\left\{ -\sum_{n>0}{1\over n}(\PINT pn)x^n \right\} x^{-1},\\
f_{++}(x) 
&\!\!\!=\!\!\!&
\exp
\left\{ -\sum_{n>0}{1\over n}{\QINT tn \over \QINT qn} (\PINT pn)x^n \right\} 
x^\beta,
\ea
and $f_{--}(x) = \omega\cdot f_{++}(x)$.\\
%
\indent
The relation between $\Psi(z)$ and 
the $q$-Virasoro generators $\Lambda^\pm(z)$ 
is simply written as 
\be
[\,\Lambda^\pm(z),\Psi(w)\,] =
\mp p^{\mp{1\over 2}} (\QINT p1)
\delta\left({w\over z}\right)\,:\!\Lambda^\pm(w)\Psi(w)\!:.
\ee


As we shall see explicitly in the next subsection,
in the limit of
$q$ and $t$ tend to $0$ with $p$ and $t^{-{|n|\over2}} h_n$ fixed,
the relations \eq{eq:ElliPP}--\eq{eq:ElliSS2}
reduce to those of $k=1$ $U_q(\widehat{sl}_2)$.
Therefore, the algebra generated by 
$S^\pm(z)$ and $\Psi(z)$ can be regarded 
as an elliptic generalization of $U_q(\widehat{sl}_2)$ with level-one.
We can regard $p$, $q$ and $t$ 
as three independent parameters. 
Even in this case, 
screening currents \eq{eq:SC21} and \eq{eq:SC22} and new currents 
$\Psi_\pm(z) \equiv 
\,:\!S_+(p^{\pm{1\over4}}z)\, S_-(p^{\mp{1\over4}}z)\!:$ 
generate an elliptic algebra. 
These extended algebras may help us to investigate 
elliptic-type integrable models.


In the sense of analytic continuation,
these relations are also rewritten by 
using elliptic theta functions\cite{rFFr},
\be
S_\pm(z) S_\pm(w) 
=
U_\pm\left({w\over z}\right) 
S_\pm(w) S_\pm(z),
\label{eq:ElliSS12}
\ee
with
\be%
U_\pm(x) =
-x^{1-2\beta}
\exp\left\{ \sum_{n\neq0}{1\over n} 
{ q^{n\over2} t^{-n} -q^{-{n\over2}} t^n \over q^{n\over2}-q^{-{n\over2}} }
x^n \right\}
=
-x^{2(1-\beta)} {\vartheta_1(px;q)\over\vartheta_1(px^{-1};q)}, 
\ee
and $U_-(x) = \omega\cdot U_+(x)$.
Note that $U_\pm(x)$ are quasi-periodic functions, namely for
$U_+(x)$, we have
\be
U_+(qx) = U_+(x), \qquad
U_+(e^{2\pi i}x) = e^{-4\pi i\beta} U_+(x).
\ee 

It should be noted that 
the screening currents of $q$-${\cal W}$ 
algebra \cite{rFFr,rAKOS} and
$U_q(\widehat{sl}_N)$ \cite{rAOS} also obey similar elliptic relations.


\subsubsection{two $c=1$ limits}


It is possible to consider two different $c=1$ limits of \v.
They are related with the (quantum) affine algebra of
$A_1^{(1)}$-type.
\\

\noindent(A) Let us consider the limit,
$q \to 0 $, $t \to 0$, $p$ is fixed. 
In this limit we must have $\beta \to 1$.
thus, this is a ``$c=1$'' limit (see (\ref{e:a9})).
We will see that the screening currents $S_\pm(z)$ 
reduces to 
the Frenkel-Jing realization of $U_{q}(\widehat{ sl}_2)$ at 
level-one \cite{rFJ}.

Introducing rescaled bosons as 
\begin{eqnarray}
a_n =-h_n q^{|n|/2}p^{-|n|/4} [2n]\qquad (n\neq 0),\qquad 
a_0=-2 h_0,\qquad
Q = -2 Q_h,
\end{eqnarray}
we obtain
\ba
&&[a_n , a_m]= \frac{[2n][n]}{n} \delta_{n+m,0} , \quad 
[a_n , Q]= 2\delta_{n,0}, \\
&&{S}_{-}(z) \rightarrow\exp\left\{\sum_{n=1}^{\infty} 
\frac{1}{[n]} a_{-n} z^n p^{-\frac{n}{4}}\right\}
\exp\left\{-\sum_{n=1}^{\infty} 
\frac{1}{[n]} a_n z^{-n} p^{-\frac{n}{4}}\right\}
e^{Q}z^{a_0}, 
\cr
&&{S}_{+}(z)\rightarrow \exp\left\{-\sum_{n=1}^{\infty} 
\frac{1}{[n]} a_{-n} z^n p^{\frac{n}{4}}\right\}
\exp\left\{\sum_{n=1}^{\infty} 
\frac{1}{[n]} a_n z^{-n} p^{\frac{n}{4}}\right\}
e^{-Q}z^{-a_0},
\label{eq:FJ}
\ea
where $[n]=({p^{\frac{n}{2}}-p^{-\frac{n}{2}}})/
({p^{\frac{1}{2}}-p^{-\frac{1}{2}}})$.  After 
replacing $p^{\frac{1}{2}}\to q$,
we can identify screening currents ${S}_{\pm}$ 
with the Frenkel-Jing realization of the 
Drinfeld currents of $U_{q}(\widehat{ sl}_2)$.

It is quite unfortunate that at this limit,
the \v current becomes singular. Namely,
it seems difficult to extract nontrivial
object from $T(z)$ at this limit. Thus it is still a challenging
problem to find a Sugawara construction for $U_{q}(\widehat{ sl}_2)$.
The limit discussed in the next paragraph (B)
may be relevant to this problem.
\\


\noindent(B)
Next, we consider the limit of $\beta \to 1$ with $q$ fixed.
This is another ``$c=1$'' limit.
In this limit, the screening currents degenerate to 
the Frenkel-Kac realization of the level-one $\widehat{sl}_2$ \cite{rFK}.

Introducing rescaled bosons as 
\begin{eqnarray}
  a_n = \frac{ 2 n}{q^{\frac{n}{2}}-q^{-\frac{n}{2}}} h_n \quad
  (n \neq 0), \quad 
  a_0 = 2h_0,   \quad
  Q =  2 Q_h,
\end{eqnarray}
we obtain
\ba
&&[a_n , a_m]= 2 n \delta_{n+m,0} , \quad 
[a_n , Q]= 2\delta_{n,0}, \\
&&{S}_{\pm}(z) \rightarrow\exp\left\{ \pm \sum_{n\neq 0}
\frac{a_{n}}{n}  z^{-n} \right\}
e^{\pm Q}z^{\pm a_0}.
\ea
{}From these, it can be seen that the screening currents have
reduced to the Frenkel-Kac realization of $\widehat{sl}_2$.

In this limit, \v survives and satisfies the relation
\begin{eqnarray}
&&\frac{\sqrt{(1-q{w \over z})(1-q^{-1}{w \over z})}}{1-{w \over z}}T(z)T(w)-
T(w)T(z)\frac{\sqrt{(1-q{z \over w})(1-q^{-1}{z \over w})}}
{1-{z \over w}}\\
&=&2(1-q)(1-q^{-1})
{w \over z} \delta'\left({w \over z} \right).\nonumber
\end{eqnarray}

It is shown that all the vectors in the Fock space
spanned over the vacuum $|0\rangle$ are singular vectors of 
this algebra
by studying the Kac determinant at level-one.
This fact and the normalization of the 
bosonic oscillators in this limit mean that the 
the Fock space is 
spanned by the Schur symmetric polynomials.
This degeneration is not accidental because the Macdonald 
polynomial $P_\lambda(x;q,t)$ reduces to the 
Schur polynomial in the limit $t\rightarrow q$.

In this ``$c=1$'' limit, \v survives and acts on the Fock space
on which the bosonized currents of $U_q(\widehat{sl}_2)$ can
also act. The relationship between these algebras
has not been made clear yet.
\\


\subsection{$c=0$ topological $q$-superconformal model}


The $c=0$ Virasoro algebra
describes a topological conformal model. 
We shall call one of the screening currents of this model $G^+(z)$,
and set $G^-(z)=:(G^+(z))^{-1}:$.
The screening charge $\oint dz G^+(z)$ plays the role of the BRST
operator.
Since the model is topological, 
the energy-momentum tensor $L_{c=0}(z)$ should be BRST 
exact. Actually, we have 
$\left\{ \oint dz' G^+(z') , G^-(z) \right\} = 2 L_{c=0}(z)$.

There exists a similar structure in the $q$-Virasoro case\footnote{
This was inspired by the discussion with T.~Kawai}.
Let us consider the case when $c=0$, {\it i.e.}, $\beta=3/2$ ($q=p^{-2}$).
Denote one of the screening current $S_+(z)$ and 
the normal ordering of its inverse
as $G^+(z)$ and $G^-(z)$, respectively
\be
G^\pm(z) =
\,:\exp\left\{
\pm\sum_{n\neq0}{h_n \over p^{n\over 2}-p^{-{n\over 2}}} z^{-n} 
       \right\}:\, 
e^{\pm2\sqrt\beta Q}
z^{\pm2\sqrt\beta h_0}.
\ee
\proclaim Proposition 6. The fields $T(z)$ and $G^\pm(z)$ satisfy 
the relations
\begin{eqnarray}
f\left( {w\over z}\right) T(z) T(w) - T(w) T(z) f\left( {z\over w}\right) 
&\!\!\! = \!\!\!& 
-(p^{1\over 2}+p^{-{1\over 2}})(p^{3\over 2}-p^{-{3\over 2}}) 
\left( \delta\left( {wp\over z}\right) 
- \delta\left( {w\over zp}\right) \right), \nonumber \\
f\left( {w\over z}\right) T(z) G^-(w) - G^-(w) T(z) f\left( {z\over w}\right) 
&\!\!\! = \!\!\!&  G^-(z) (p^{3\over 2}-p^{-{3\over 2}}) 
\left( p^{2}\delta\left( {wp\over z}\right) 
-p^{-2}\delta\left( {w\over zp}\right)\right), \nonumber \\
\left\{\oint dz G^+(z), G^-(w) \right\} &\!\!\! = \!\!\!& 
{p^{-1}\over w^2 (p-p^{-1})(p^{1\over 2}-p^{-{1\over 2}})}
\left(  T(w) - (p^{1\over 2}+p^{-{1\over 2}}) \right),  \nonumber\\
\left[ T(z), \oint dw G^+(w) \right]  
&\!\!\! = \!\!\!&   0 , \qquad\quad
\left\{ G^\pm(z),G^\pm(w)\right\}  =  0.
\label{eq:qTopological}
\end{eqnarray}
where $f(x)$ is given by (\ref{structure}) with $q=p^{-2}$.

\noindent
Note that $G^-(z)$ is a primary field and 
its commutation relation with $T(z)$ is given by the same function $f(x)$
in the defining relation of the $q$-Virasoro algebra.

We can regard these relations as a $c=0$ topological $q$-Virasoro algebra.
The screening charge $\oint dz G_+(z)$ may play the role of BRST operator
which reduces the bosonic Fock space to 
irreducible representation space of the $q$-Virasoro algebra.
At the value of the coupling constant $\beta=3/2$,
the central charge vanishes and 
the entire Fock space contains only BRST trivial states,
except for the vacuum state. 
The $q$-Virasoro generator itself (up to a constant)
is given by a BRST exact form. 
Thus the $\beta=3/2$ $q$-Virasoro algebra is a topological field theory
same as $q=1$ case.

The relations between the currents $G^\pm(z)$ and $\Lambda^\pm(z)$ are
written as
\begin{eqnarray*}
f\left( {w\over z}\right) \La^\pm(zp^{\pm1}) G^+(w) 
- G^+(w) \La^\pm(zp^{\pm1}) f\left( {z\over w}\right) 
&\!\!\!=\!\!\!& \mp p^{\mp1}(p^{3\over 2}-p^{-{3\over 2}})
\delta\left( {w\over z}p^{\mp1}\right) G^+(z) ,\\
f\left( {w\over z}\right) \La^\pm(z) G^-(w) 
- G^-(w) \La^\pm(z) f\left( {z\over w}\right) 
&\!\!\!=\!\!\!& \pm p^{\pm2}(p^{3\over 2}-p^{-{3\over 2}})
\delta\left( {w\over z}p^{\pm1}\right) G^-(z) ,
\end{eqnarray*}
\vspace{-7mm}
\begin{eqnarray}
\left\{G^+(z), G^-(w) \right\} 
&=&{p^{-3}\over zw^2 (p-p^{-1})(p^{1\over 2}-p^{-{1\over 2}})} \\
&\times&
\left(  \La^+\left( {w}\right)\delta\left( {w\over zp}\right) 
+ \La^-\left( {w}\right)\delta\left( {wp\over z}\right)
-\left( p^{1\over 2}+p^{-{1\over 2}}\right)
\delta\left( {w\over z}\right) \right). \nonumber
\end{eqnarray}
Their Fourier modes given by
$G^\pm(z) = \sum_n G^\pm_n z^{-n}$ and 
$\La^\pm(z) = \sum_n \La^\pm_n z^{-n}$
satisfy 
$$
\left\{G^+_{n+1}, G^-_{m+1} \right\} =
{p^{-3}\over (p-p^{-1})(p^{1\over 2}-p^{-{1\over 2}})}
\left(  \La^+_{n+m-1} p^{-n} + \La^-_{n+m-1} p^n 
-\left( p^{1\over 2}+p^{-{1\over 2}}\right)\delta_{n+m-1,0} \right).
$$
Note that, 
the relation between $G^+$ and $G^-$ can be 
expressed in other ways.
For example, use the fact that 
$ G^+(z) G^-(w) {z/w} + G^-(w) G^+(z) {w/z} $ and
$\left\{ G^+(z), G^-(w) \right\} z^{-r}{z^2 w^2/(z+w)}$
are also written by $\La^\pm$ for any $r\in\bC$.

In the $q=1$ case, $c=0$ topological algebra 
can be constructed from 
$N=2$ superconformal algebra by the 
operation so-called ``twisting''\cite{rEY}.
What we have obtained here is a deformation of this twisted 
superconformal algebra. So far, 
we have not been able to find a mechanism
of ``untwisting'' in the deformed case.
We expect that our topological $q$-Virasoro algebra 
helps us to find a 
supersymmetric generalization of the $q$-Virasoro algebra \v
and a deformed twisting operation.


\subsection{$c=-2$ \v and $c=1$ $\Winf$ algebra}


Since $c=0$ Virasoro algebra is realized by the differential operator 
$L_n = - z^n D$ with $D = z \partial_z$,
one can expect that $q$-Virasoro algebra \v has a similar 
representation by the difference 
or shift operator as
$T_n \sim z^n q^D$.
However, this shift operator is nothing but the generating function of 
the $c=0$ $\Winf$ generators.
Thus we expect some relations between the $q$-Virasoro and the $\Winf$ algebra.
Indeed this is the case when $\beta=2$, 
we show a relation with $c=1$ $\Winf$ algebra.

First, recall that the $q$-Virasoro generator is expressed by 
the fundamental vertex operator and its dual $V_{2,1}^\pm(z)$ 
as eq.\ \eq{eq:TSbyV21}.
When $\beta=2$, {\it i.e.}, $c=-2$, ($q=1/p$), 
they reduce to the fermions such that
\be
\{V_{2,1}^+  (z),V_{2,1}^-  (w)\}
={1\over z}\delta\left( {w\over z}\right),\qquad
\{V_{2,1}^\pm(z),V_{2,1}^\pm(w)\}=0.
\ee
On the other hand, the generating function of $c=1$ $\Winf$ algebra
is known to be also realized by a complex fermion. 
Therefore, we have found;

\proclaim Proposition 7.
The $\beta=2$ ($c=-2$) $q$-Virasoro algebra $T(z) = \La^+(z) + \La^-(z)$ 
generates the $c=1$ $\Winf$ algebra
and it is realized by fermions $V^\pm_{2,1}(z)$ as follows;
\be
\La^\pm(zq^{{1\over 2}\mp1}) =
\,:V_{2,1}^-(z) q^{\mp D} V_{2,1}^+(z):\, q^{\mp{1\over 2}}.
\label{eq:TbyFermion}
\ee


Next we show the relation between $q$-Virasoro and $\Winf$ algebras
more explicitly comparing their commutation relations.
Let
\be
X^k(z) \equiv
\,:\exp\left\{\sum_{n\neq0}{1-q^{kn}\over 1-q^n} h_n z^{-n} \right\}:\, 
q^{k\sqrt 2 h_0}
\equiv
\sum_{n\in\bZ} X^k_n z^{-n}.
\ee
Note that the $q$-Virasoro generator is now $T(z)=X^1(z) + X^{-1}(z)$.
Then
\ba\nn
\left[\, X^k(z),X^\ell(w) \,\right] \&
= {(q^k-1)(q^\ell-1)\over q^{k+\ell}-1}
\left( X^{k+\ell}(z) \delta\left( q^k{w\over z}\right) 
- X^{k+\ell}(w) \delta\left( q^\ell{z\over w}\right)\right), 
\cr
\left[\, X^k(z),X^{-k}(w) \,\right] \&
= (q^{k\over 2}-q^{-{k\over 2}})^2 \left(  
\left( 1+\sum_{n\neq0}{1-q^{kn}\over 1-q^n} 
n h_n z^{-n}\right) \delta\left( q^k{w\over z}\right) 
-\delta'\left( q^k{w\over z}\right) 
\right),
\ea
for $k,\ell,n,m\in\bZ$.
Their modes
\be
W^k_n = {X^k_n\over q^k-1} - {1\over 1-q^{-k}} \delta_{n,0},\qquad
W^0_n = {n\over q^n-1} h_n,
\ee
for $k\neq0$, satisfy
\ba
\left[\, W^k_n,W^\ell_m \,\right] \&
= (q^{-km}-q^{-\ell n}) W^{k+\ell}_{n+m} + 
  {q^{-km}-q^{-\ell n}\over 1-q^{-k-\ell}} \delta_{n+m,0},
\cr
\left[\, W^k_n,W^{-k}_m \,\right] \&
= (q^{-km}-q^{kn}) W^0_{n+m} + n q^{kn}\delta_{n+m,0}.
\label{eq:Winfty}
\ea
This is nothing but the algebra of 
the generating functions for the $c=1$ $\Winf$ generators
$W^k_n = W\left( z^nq^{-kD}\right)$
in the notation of \cite{rAFMO}.

The $c=0$ $\Winf$ algebra has a meaning of an area-preserving diffeomorphism
and relates with classical membrane.
We expect this relation 
between the $q$-Virasoro algebra with the $\Winf$ algebra 
is a key for
a geometrical interpretation of \v and
the quantization of the membrane.

%
%

\section{Summary and further issues} 


Our presentation has aimed to show that
the new Virasoro-type elliptic algebra \v defined by eq.\ \eq{e:a1.2}
can be regarded as a universal symmetry of the massive integrable models.


The algebra \v has two parameters $p$ and $q$ ($qp^{-1}=q^\beta$)
and it can be regarded as a generating function for 
several different Virasoro-type symmetry algebras
appearing in solvable models.
%
%
At the limit of $q\rightarrow 1$, the algebra \v 
reduces to the ordinary Virasoro algebra 
with the central charge $c$ related to $\beta$ \eq{e:a8}.
When we consider the limit of $q\rightarrow p$, 
it reduces to the  
$q$-Virasoro algebra of Frenkel-Reshetikhin \eq{qvirFR}.
When $q\rightarrow 0$, 
we obtain Jing's generating operators for
the Hall-Littlewood polynomials \eq{eq:BosonHall}.
The topological algebra \eq{eq:qTopological} and 
the $c=1$ $\Winf$ algebra \eq{eq:Winfty} are constructed from 
the special cases $p=q^{-1/2}$ and $p=1/q$ respectively.

One of the peculiar features of the algebra \v is its non-linearity.
Since it is a quadratic algebra like the Yang-Baxter relation 
for the transfer matrix,
the associativity is quite non-trivial \eq{YB-for-T},
and the Yang-Baxter equation 
determines the structure function uniquely, {\it i.e.},
fixes the algebra itself!
Moreover it turns out to be the Zamolodchikov-Faddeev algebra 
of the particle-creation operators for the XYZ and 
the sine-Gordon models \eq{eq:ZFXYZ}.

The next essential nature is its infinite-dimensionality, 
which connects with the integrability of massive models.
The representation space of the algebra \v possesses rich structure 
enough to describe the physical space of massive models.
Despite its non-linearity, 
the Kac determinant is very similar to that of the Virasoro case 
\eq{Kacconj}. 


The algebra \v is realized by the free fields 
\eq{eq:qVirFFR1} and \eq{eq:qVirFFR2}
in a quite simple way.
It shows us that the integrability of the model due to \v symmetry
can be investigated in a natural manner in terms of the 
bosonic field.
Furthermore this free field realization is described by 
a deformed Miura transformation \eq{eq:qMiura},
and the deformed Miura transformation 
brings about an interesting analogy with the dressed vacuum form.
A generalization of this transformation gives
the $q$-deformed $\cal W$ algebra \cite{rFFr,rAKOS}.

To study infinite dimensional algebras, {\it e.g.},
not only Virasoro and Kac-Moody algebras but also \v,
one of the most essential oubjects in the representation theory is 
the screening current. 
Using the screening charge, one can
define the physical states by the BRST method \eq{eq:BRST},
write down the null states \eq{eq:qVirSing}, 
and study the nontrivial monodromic property of the 
screened vertex operators. 
The null states  
relate with the wave functions of the 
Ruijsenaars model \eq{eq:MacOp} which is  
a relativistic generalization of the Calogero-Sutherland model. 
The Hamiltonian of this model
is realized by the positive modes of the \v generators \eq{e:b7.3}
and causes the correspondence between the null states and excited states.
The monodromy matrix is connected to the ABF model.
The exchange relation of the vertex operators
possesses a quantum group structure 
characterized by the solution of the face type 
Yang-Baxter equation\eq{eq:ExchangeABF}
and leads to the identification of the deformed Virasoro vertex operators
with that of the ABF model.
Moreover the screening currents show us an hidden symmetry;
that is an elliptic generalizaton of 
$k=1$ $U_q(\widehat{sl}_2)$ algebra (Prop. 5).
\\


Finally, we mention some comments on further issues 
coming from mathematical or physical points of view.

Mathematically, there are many things to be clarified.
To obtain a fusion of the ABF model or to find a suitable primary fields,
one needs a tensor product representation or 
suitable adjoint action of the algebra \v.
In other words, a co-product structure must be discovered. 
Since the algebra, however, is quadratic, it seems a highly non-trivial task.
Studying the limit $q\rightarrow 0$ may help us to reveal it.

Dispense with a help of free fields,
the correlation function must be determined 
by a difference equation coming from the Ward-Takahashi identity.
To find this equation, 
we need to recognize a geometrical meaning of the algebra \v;
how is a difference operator realized in \v?
Restricting the $q$-${\cal W}_{1+\infty}$ algebra \cite{rKLR},
which is defined as an algebra of higher pseud-difference operators,
to the first order difference one,
we might obtain the algebra \v and its difference operator realization.
Seeking for a realization on the infinite $q$-wedge \cite{rKMS}
might be able to connect both problems mentioned above; 
a co-product and a geometrical interpretation.

The relation with the quantum affine Lie algebras also 
seems to be of some interest.
Frenkel-Reshetikhin's $q$-Virasoro algebra was constructed 
by a $q$-Sugawara method \cite{rFR} from 
the $U_q(\widehat{sl}_n)$ at the critical level.
Are there any such constructions for the algebra \v from 
$U_q(\widehat{sl}_n)$ or, more hopefully, from 
their elliptic generalization ${\cal A}_{q,p}$?


Physically, we anticipate many applications.
There are two approaches to investigate the 2-dimensional integrable models, 
one is an Abelian method based on the algebraic Bethe-ansatz
and the other is a non-Abelian one based on
the Virasoro algebra and its generalizations.
The latter approach describes the model more in detail,
however its applicability is restricted 
only to critical phenomena and some off-critical 
trigonometric-type models.
The deformed Virasoro algebra \v 
should be a synthesis of the massive integrable models
including elliptic-type models.

In the dual resonance model, which was a precursor of string theory,
the Veneziano amplitude has its generalizations
to non-linearly rising Regge trajectories (see for example, \cite{rFN}).
However the absence of their operator representations
has disturbed their development, including to prove a no-ghost theorem. 
In a special case, the amplitude reduces to $q$-beta function, 
which is very similar to a four-point function of our algebra \v.
We hope that \v gives 
an operator representation of a generalized Veneziano amplitude 
and opens further avenues to the exploration of new string theories.

The $c=0$ $\Winf$ algebra, an area-preserving diffeomorphism,
is a symmetry of the classical membrane.
The relation between the $c=-2$ \v algebra and $c=1$ $\Winf$ algebra 
may be a key for
a geometrical interpretation of the algebra \v and
the quantization of the membrane 
as the basic object of the $11(12)$-dimensional M(F) theory.
On the other hand, 
quantum membrane can be represented by a large $N$ matrix model \cite{rDHN}, 
and the partition functions of more general conformal matrix models
are described by eigenstates of the Calogero-Sutherland models \cite{rAMOS}.
Do the algebra \v or eigenstates of the Ruijsenaars model 
relate with a relativistic generalization of the quantum membrane? 

What is a field theoretical interpretation of the algebra \v?
The low energy 4-dimensional $N=2$ super YM theories 
are described by some integrable models, 
periodic Toda chain or elliptic CS model \cite{rGKMMM}.
If we generalize them to 5D's one by Kaluza-Klein method
with an extra dimension compactified to a circle,
then they are described by the relativistic generalization of 4D's one
\cite{rN}.
It may suggest that the Kaluza-Klein with a radius $R$ can lead to 
a relativistic generalization with the speed of light $1/R$
or a $q$-deformation with $q=e^R$ of a original theory.
Is the algebra \v understood as a Kaluza-Klein from CFT?

\vskip 5mm

\noindent{\bf Acknowledgments:}


We would like to thank 
T.~Eguchi, D.~Fairlie, B.~Feigin, E.~Frenkel,
J.~Harvey,  T.~Hayashi, T.~Inami, M.~Jimbo, 
S.~Kato, T.~Kawai,  B.~Khesin, A.~Kuniba, 
E.~Martinec, Y.~Matsuo, T.~Miwa, K.~Nagatomo, N.~Nekrasov, 
P.~Wiegmann and Y.~Yamada
for discussions and encouragements.
This work is supported in part by Grant-in-Aid for Scientific
Research from Ministry of Science and Culture.


\end{document}